**Recent Advances on Transition-Metal-Based Layered Double Hydroxides Nanosheets for Electrocatalytic Energy Conversion**

*Yuchen Wang,† Man Zhang,† Yaoyu Liu, Zhikeng Zheng, Biying Liu, Meng Chen, Guoqing Guan,* Kai Yan**

Dr. Y. Wang, M. Zhang, Y. Liu, Z. Zheng, B. Liu, Prof. K. Yan
Guangdong Provincial Key Laboratory of Environmental Pollution Control and Remediation Technology, School of Environmental Science and Engineering, Sun Yat-sen University, Guangzhou 510275, China
E-mail: yank9@mail.sysu.edu.cn

Dr. M. Chen, Prof. G. Guan
Energy Conversion Engineering Laboratory, Institute of Regional Innovation (IRI), Hirosaki University, 3-Bunkyocho, Hirosaki, 036-8561, Japan
Email: guan@hirosaki-u.ac.jp
† These authors contributed equally to this work.

**Abstract:** Transition-metal-based layered double hydroxides (TM-LDHs) nanosheets are promising electrocatalysts in the renewable electrochemical energy conversion system, which have been regarded as alternatives of noble metal-based materials. In this review, recent advances on effective and facile strategies to rationally design TM-LDHs nanosheets as electrocatalysts, such as increasing the number of active sties, improving the utilization of active sites (atomic-scale catalysts), modulating the electron configurations, and controlling the lattice facets, are summarized and compared. Then, the utilization of these fabricated TM-LDHs nanosheets for oxygen evolution reaction, hydrogen evolution reaction, urea oxidation reaction, nitrogen reduction reaction, small molecule oxidations and biomass derivatives upgrading are articulated through systematically discussing the corresponding fundamental design principles and reaction mechanism. Finally, the existing challenges in increasing the density of catalytically active sites and future prospects of TM-LDHs nanosheets based electrocatalysts in each application are also commented.

## 1. Introduction

An increasing demand of energy and high-value-added chemicals for the growth and development of modern society has inevitably accelerated the consumption of limited non-renewable fossil resources.[1] The excessive carbon emission from fossil fuels has caused adverse changes to the climate, especially global warming issue.[2] At this moment, renewable energy based technologies are urgently needed to achieve a sustainable development of human society (Figure 1).[3] Among manifold frontal technologies, those renewable energy powered electrocatalytic conversion techniques including water electrolysis and biomass upgrading have drawn tremendous interests owing to their high efficiency and non-pollution.[4] Accordingly, the architecture of high-performance electrocatalyst is essential to the efficient utilization of green energy and production of valuable chemicals.

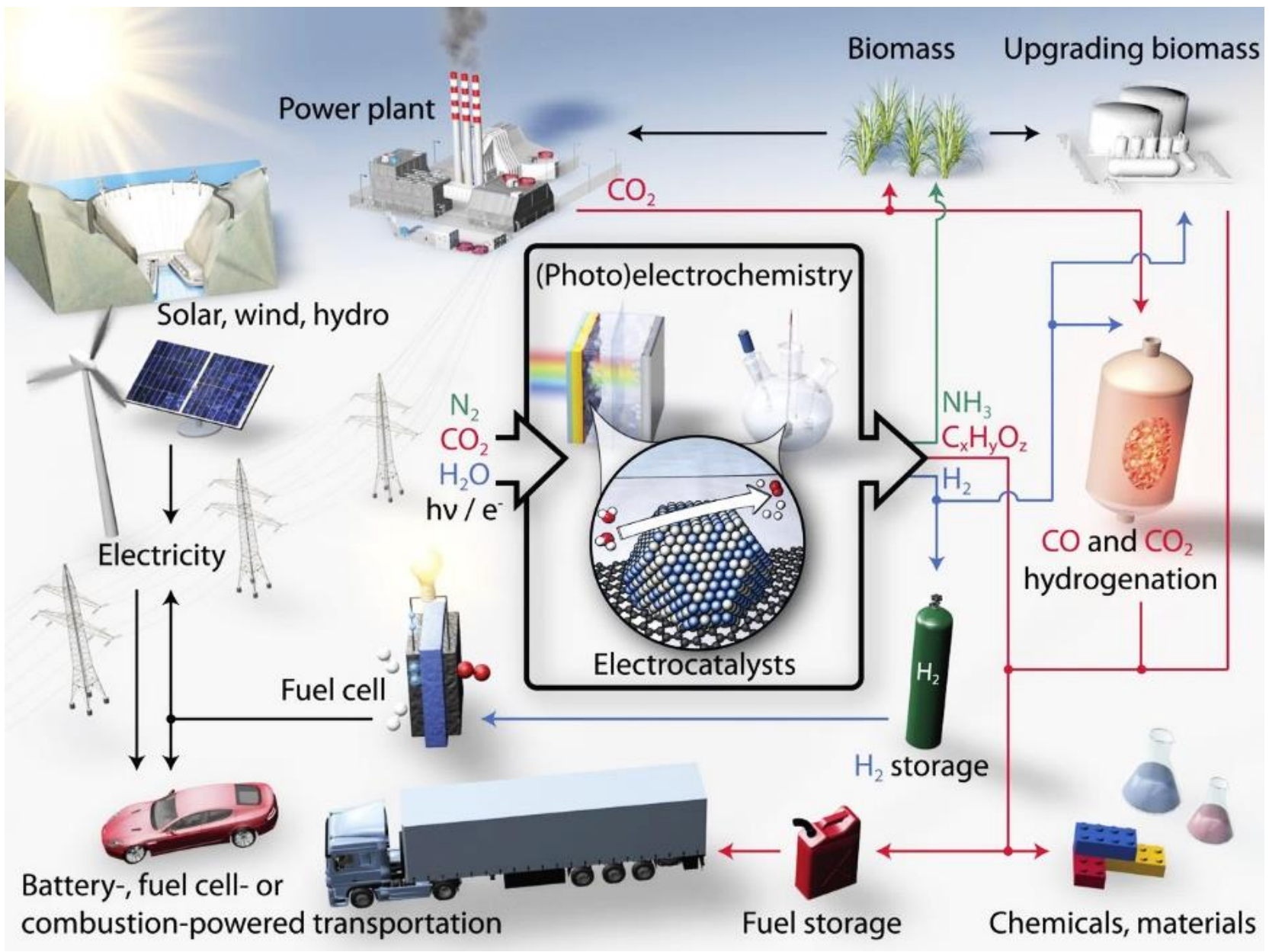

**Figure 1.** Schematic of a renewable energy landscape related to electrocatalysis. Reproduced with permission.[3a] Copyright 2017, American Association for the Advancement of Science.

Two-dimensional materials are considered as desired alternatives to noble metal-based electrocatalysts because of their unique electronic and anisotropic properties.[5] Transition-metal-based layered double hydroxides (TM-LDHs) are classical 2D electrocatalysts with earth abundance, cheapness, low toxicity and high intrinsic electrocatalytic activity.[6] As displayed in Figure 2, TM-LDHs possess a lamellar structure with the charge balance between transition metal divalent/trivalent cations ($M^{2+}/M^{3+}$) in host layers and anions ($A^{n-}$) in interlayers.[7] A versatile formula of TM-LDHs is expressed as $[M^{2+}_{1-x}M^{3+}_{x}(OH)_2]^{x+}[(A^{n-})_{x/n}\cdot yH_2O]^{x-}$, where x can be valued in the range of 0.20−0.33 [8]. As such, the adjustment of interlayer spacing and the modulation of electronic structure can be realized by tuning transition metal divalent/trivalent cations and anions in TM-LDHs.

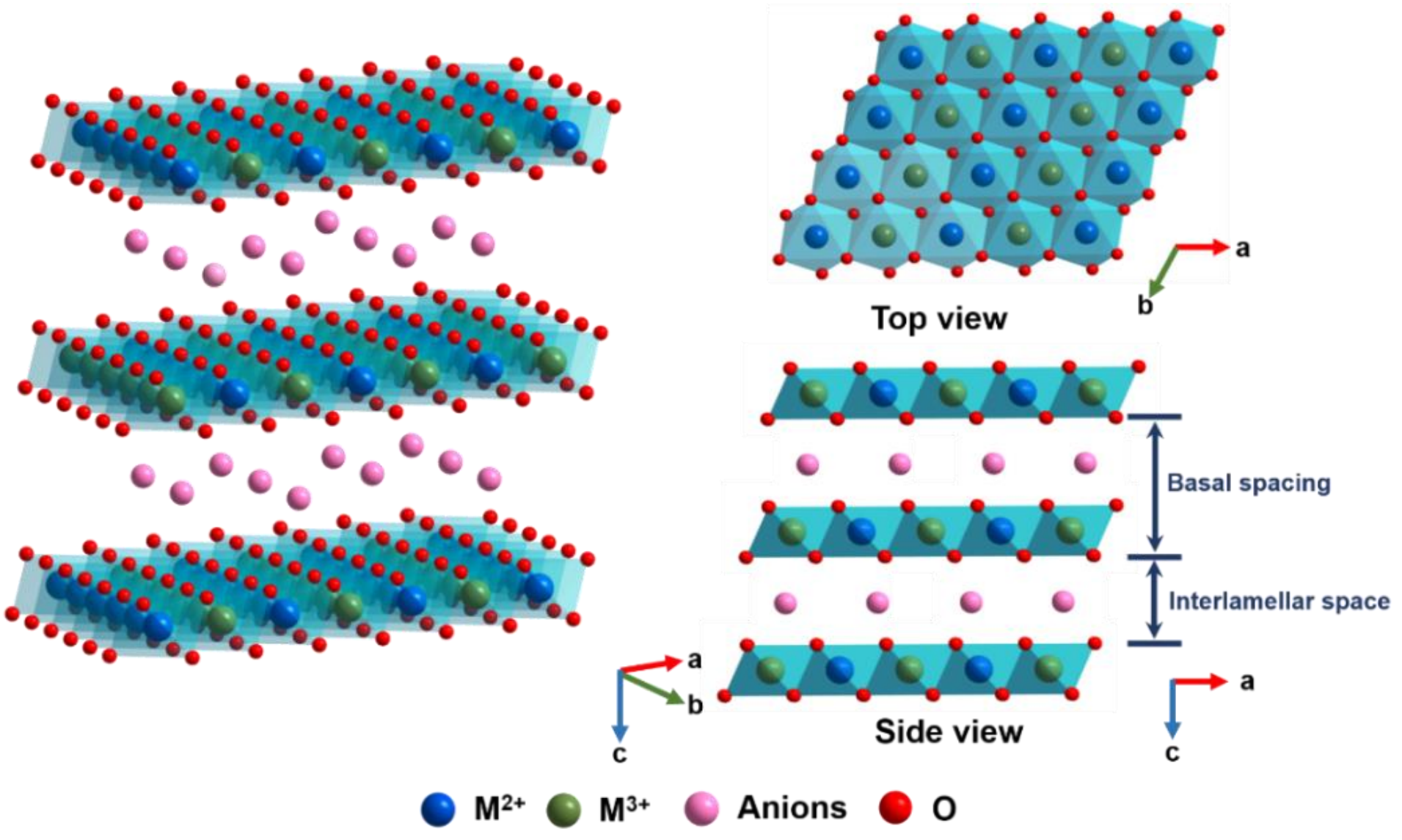

**Figure 2.** Schematic crystalline structure of TM-LDHs.

Although TM-LDHs have been intensively studied in the electrocatalysis field, insufficient active sites and limited electronic/ionic transport are main obstacles for their industrial applications.[9] To effectively overcome these shortcomings, realizing TM-LDHs nanosheets with nanometer-level thickness and reducing lateral size are necessary.[10] At the same time, it is reported that abundant defects are always generated in the prepared TM-LDHs nanosheets, which are imperative for improving the intrinsic activity.[11] Moreover, considering the lack of interlayer anions, TM-LDHs nanosheets with positively charge are prone to hybrid with negatively charged nanostructured materials (e.g., graphene, carbon nanotubes, metal oxides).[12] In addition, the electrostatic strengthened hybrid structure is beneficial for alleviating the agglomeration of TM-LDHs nanosheets, enhancing the overall conductivity and ameliorating the electronic structure, giving rise to the optimized electrocatalytic performance.[13]

In this review, a brief overview of recent advances in the development and utilization of TM-LDHs nanosheets for electrocatalytic energy conversion is presented (Figure 3). It is proposed that TM-LDHs nanosheets could be a promising category of energy materials in various electrocatalytic reactions including water splitting, urea oxidation, nitrogen reduction reaction, small molecule oxidation and biomass derivatives upgrading. Herein, state-of-the-art fabrication methodologies of TM-LDHs nanosheets are first given and followed by a summary of the latest theoretical studies and experimental progress of TM-LDHs nanosheets for electrocatalytic energy conversion. Especially, the applications of TM-LDHs nanosheets in water splitting, biomass monomers upgrading are focused. In the end, the challenges and corresponding future perspectives of TM-LDHs nanosheets as high-performance electrocatalysts are proposed.

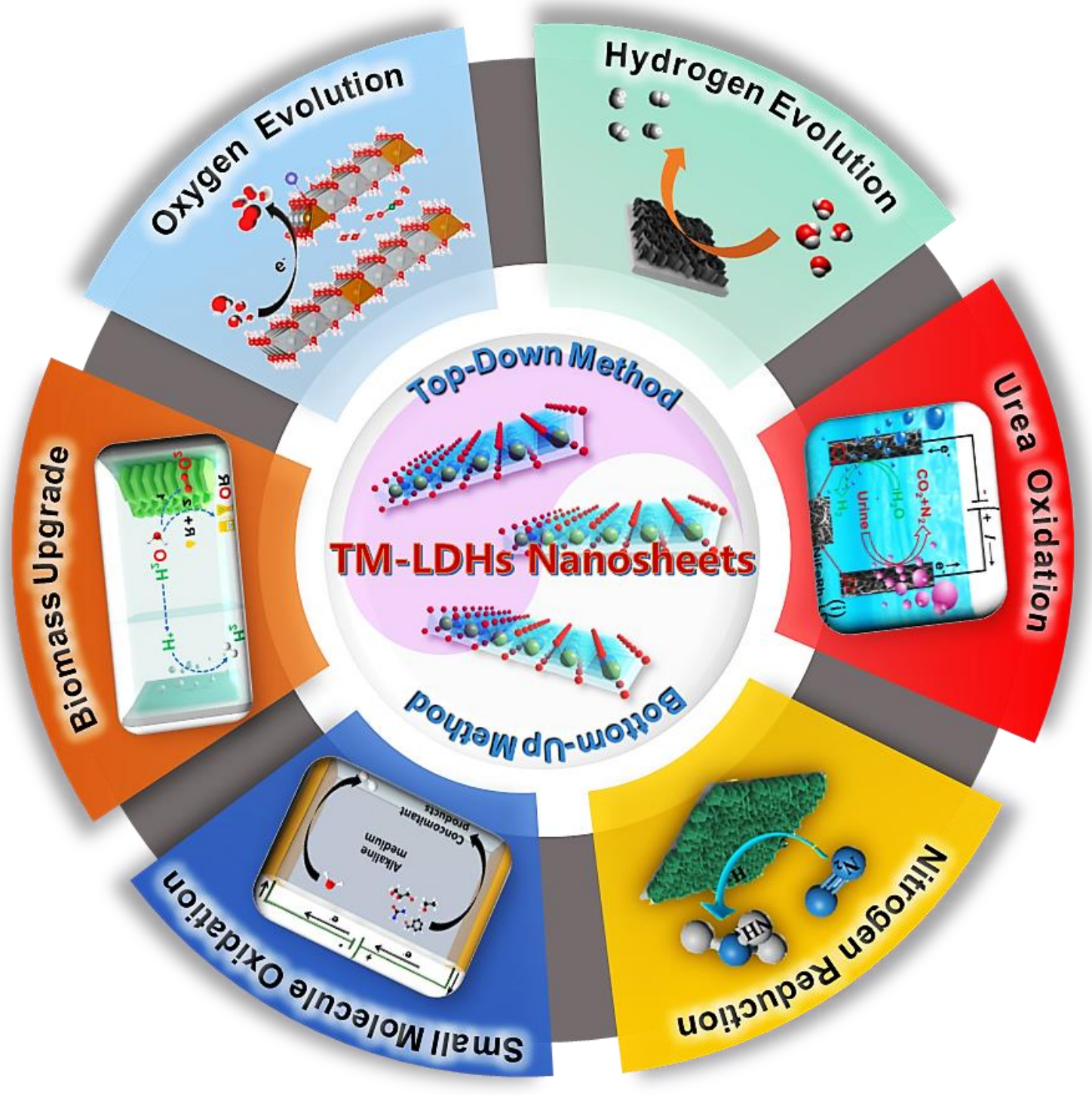

**Figure 3.** Schematic representation of synthetic strategies and electrocatalytic applications of TM-LDHs nanosheets.

## 2. Synthetic Strategies for TM-LDHs Nanosheets

In recent years, enormous synthetic strategies have been developed for fabricating TM-LDHs nanosheets, which are generally classified into top-down and bottom-up strategies.[14] The top-down strategy is typically referred as exfoliation or delamination of conventional bulk TM-LDHs to produce nanosheet structure using physical and chemical means. On the contrary, the bottom-up strategy is a direct preparation route of TM-LDHs nanosheets without pre-synthesizing bulk precursors. In this section, all relevant progresses on these two strategies are reviewed.

**2.1. Top-Down Strategy**

Up to date, the top-down strategy has been mostly used to achieve TM-LDHs nanosheets by breaking the interlayer bonds in bulk parent crystal. In 1999, Adachi-Pagano *et al.* achieved the first successful exfoliation of TM-LDHs, in which the parent ZnAl-LDHs were initially modified via the intercalation of dodecyl sulfate surfactants and then dispersed in butanol for 16 h.[15] As such, the exfoliated ZnAl-LDHs nanosheets could maintain stable for more than 8 months. It is also reported that other solvents (e.g., water, methanol, ethanol, propanol or hexane) only partially delaminated ZnAl-LDHs, revealing the importance of solvent type on the stripping effect.

Among various solvents, formamide has been considered as a general exfoliation agent for TM-LDHs.[16] Liu *et al.* proposed a universal exfoliation mechanism of TM-LDHs in formamide based on experimental results (Figure 4a), in which TM-LDHs firstly undergo a quick swelling process to weaken the strong hydrogen bonds between carbonyl groups of formamide and hydroxyl groups of TM-LDHs slabs, and as such, the subsequent exfoliation process can proceed under continuous mechanical treatments to obtain TM-LDHs nanosheets.[17] Inspired by this work, myriad mono- or few-layer TM-LDHs (e.g., NiCo-, NiFe-, NiV-, NiAl-, CoFe-LDHs) nanosheets and their heterostructures have been successfully obtained on the basis of such a delamination approach.[10b,18]

Apart from using poisonous polar organic solvents, water is also utilized as a mild solvent for liquid exfoliation. For example, Wei *et al.* introduced sodium hydroxide/urea aqueous solution to derive ZnAl-LDHs nanosheets with a thickness of 0.6 nm at −10 °C.[19] As seen in Figure 4b, the insertion of sodium hydroxide/urea hydrates can destroy the original hydrogen bond network of Zn-Al LDHs and then promote the rapid delamination process at a subzero temperature. Lately, Chen *et al.* applied a novel hydrothermal approach to achieve NiFe-LDHs nanosheets based on the Ostwald ripening phenomenon.[20] Herein, compared with other liquid exfoliation strategies, bulk NiFe-LDHs with different sizes are firstly grown on the copper mesh and then delaminated via a dissolve-redeposition process without using any surfactant and organic solvent. After an optimal hydrothermal period of 8 h, NiFe-LDHs nanosheets with a thickness of 7−9 nm and a lateral size of 1 μm can be ultimately collected. Although above methods have successfully used to synthesize TM-LDHs nanosheets, conventional liquid exfoliation techniques are still suffered from tediously long processing time and reaggregation of exfoliated nanosheets. Therefore, high-efficient and eco-friendly exfoliating strategies for the fabrication of stable ultrathin TM-LDHs nanosheets are urgently needed.

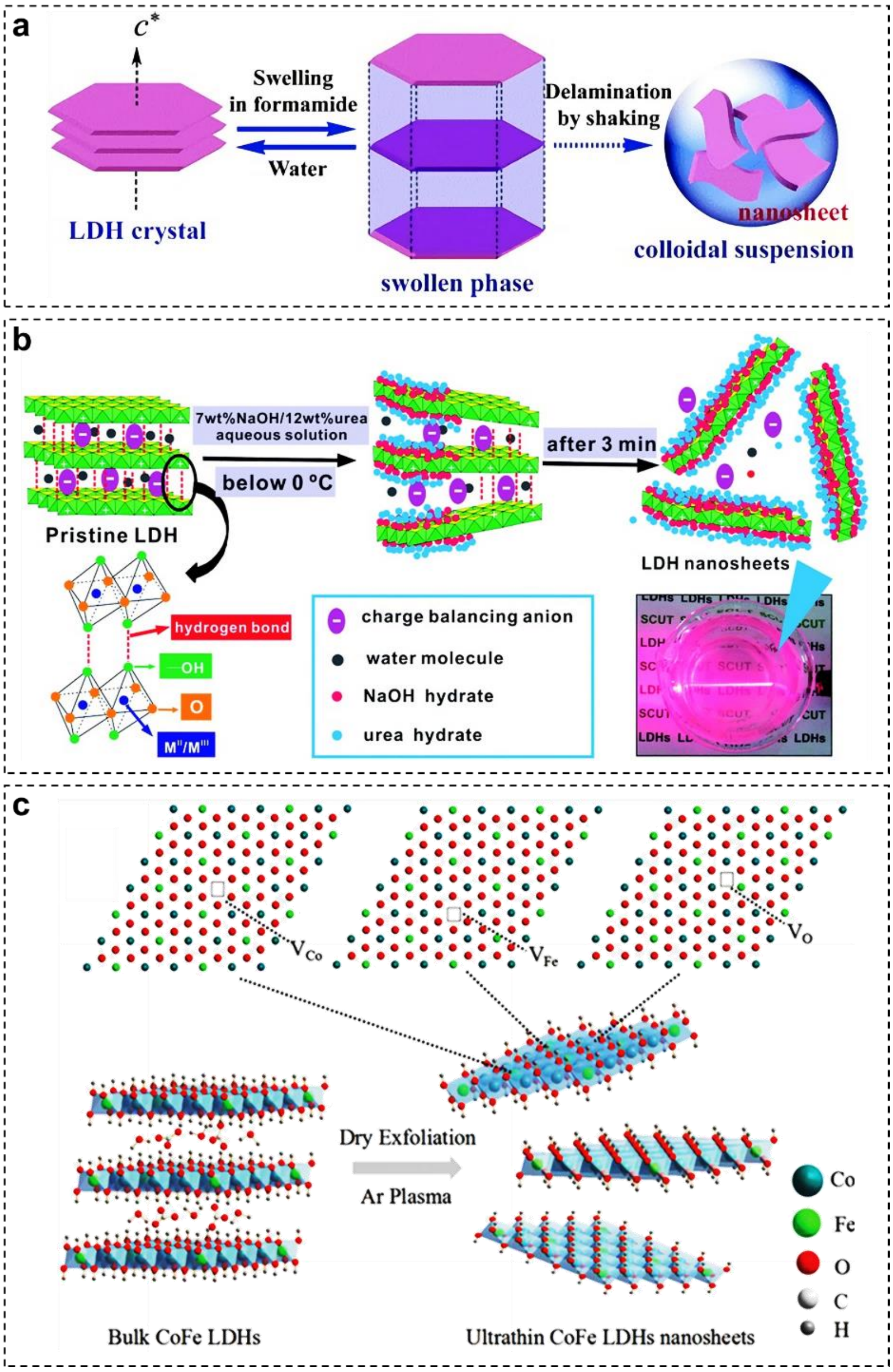

**Figure 4.** (a) Exfoliation mechanism of TM-LDHs in formamide. Reproduced with permission.[17] Copyright 2006, American Chemical Society. (b) Exfoliation of TM-LDHs in sodium hydroxide/urea aqueous solution. Reproduced with permission.[19] Copyright 2014, Royal Society of Chemistry. (c) Schematic diagram of Ar plasma exfoliation of CoFe-LDHs. Reproduced with permission.[21] Copyright 2017, Wiley.

Plasma technology has been widely used to exfoliate layered materials such as graphite,[22] transition metal oxides,[23] and also TM-LDHs.[21] For a typical plasma exfoliation of TM-LDHs, the bulk TM-LDHs can be initially synthesized through a hydrothermal method and then attacked by plasma-induced active particles under a certain atmosphere (e.g., Ar, $N_2$, $O_2$). Herein, these active particles will break the chemical interaction between host layers so that corresponding interlayer anions can separate TM-LDHs lamellar structure. Such a plasma exfoliation technique has the advantages of cleanness, time-saving and low energy consumption in comparison with the traditional liquid exfoliation technique.[24] In 2017, Wang *et al.* firstly applied Ar plasma to obtain ultra-thin CoFe-LDHs nanosheets (Figure 4c).[21] It is found that the plasma etching process not only reduced the thickness of CoFe-LDHs from 20 nm to 0.6 nm but also generated Co, Fe and O vacancies, which are advantageous to the electrocatalytic performance. Later on, Liu *et al.* successfully delaminated NiCo-LDHs via 15 s of Ar plasma treatment.[25] To expand the application of this technique, $N_2$ and water plasma have been also utilized to realize the fabrication of CoFe-LDHs and CoAl-LDHs nanosheets.[26]

### 2.2. Bottom-Up Strategy

Compared with the top-down strategy, the bottom-up strategy is relatively simple to directly synthesize TM-LDHs nanosheets from transition metal salts. To date, various bottom-up methods including reverse microemulsion method, co-precipitation method with nucleation/growth controlled inhibitor, pulsed-laser ablation and ultrasonic/mechanical stirring methods have been developed.

The reverse microemulsion approach is effective for preparing TM-LDHs nanosheets via the introduction of aqueous co-precipitation system into oil phase. Generally, the fabrication process is illustrated in Figure 5a: 1) transition metal ions and urea are mixed in micelles, which are surrounded by surfactant; 2) TM-LDHs are generated along with the slow hydrolysis of urea; 3) the formation of TM-LDHs nanosheets in a confined environment.[27] Herein, the water/surfactant molar ratio can be adjusted to control the size of TM-LDHs nanosheets. For instance, Zhao *et al*. applied the reverse microemulsion method to obtain NiTi-LDH nanosheets, which possessed a thickness of 0.9 nm with a particle size of 20 nm.[28] However, it should be noted that long reaction time and difficult removal of surfactants cannot be ignored by using this approach.

As regards the traditional co-precipitation or hydrothermal process, numerous inhibitors including formamide, ethyl alcohol and ethylene glycol are added to weaken the interactions between TM-LDHs host layers, thereby preventing their growth along c-axis. For example, Zhao *et al*. applied a co-precipitation method to obtain monolayer NiFe-LDHs precursors with a thickness of 1.0 nm and a lateral size of 30 nm in formamide.[29] The subsequent ultrasonication effectively reduced the corresponding sizes to 0.6 nm and 1.5−3.0 nm, respectively. Wang *et al*. adopted a facile hydrothermal method with the aid of ethyl alcohol to construct oxygen-deficient NiAl-LDHs nanosheets on carbon fiber paper (Figure 5b), in which the growth of NiAl-LDHs nanosheets proceeded without adding any alkaline source.[30] The obtained petal-like NiAl-LDHs nanosheets provided a high specific surface area with desired intrinsic conductivity, which is vital to achieve superior electrocatalytic performance. Li *et al*. successfully prepared stable single-layer CoAl-LDHs using a hydrothermal method in ethylene glycol.[31] It is considered that the formation of the single-layer CoAl-LDHs should be mainly attributed to the inherent chelation and absorption ability of ethylene glycol.

To avoid the use of toxic organic solvents, the electrodeposition method and mechanical strategies including laser ablation and ultrasonic/mechanical stirring methods have been extensively reported to prepare TM-LDHs nanosheets. Electrodeposition method is a facile and eco-friendly process to fabricate TM-LDHs nanosheets on suitable conductive substrates. Typically, nitrate ions in the electrolyte are reduced under constant current or potential to produce hydroxide ions, which can further combine with transition metal cations to form TM-LDHs nanosheet arrays on the surface of the substrate.[6b,32] The morphologies of as-prepared TM-LDHs nanosheets are highly dependent on electrodeposition parameters, such as reaction time and precursor concentration.[33] Wang *et al.* employed *in situ* liquid-phase transmission electron microscopy to investigate the influence of electrodeposition parameters on the growth mode and density of CoFe-LDHs nanosheets.[34] It is found that the growth of CoFe-LDHs nanosheets was rapid at the initial stage and then unchanged after reach a critical areal coverage.

Impressively, the vertical or lateral growth of these nanosheets could be modulated by the precursor concentration, providing a guidance for controllably electrochemical synthesis of TM-LDHs nanosheets.

With respect to mechanical strategies, Hunter *et al.* employed the pulse-laser ablation technique to prepare NiFe-LDHs nanosheets as follows: 1) NiFe-LDHs precursors and pure metallic powders are magnetically stirred in alkalike water (pH = 10); 2) the mixed solution is irradiated to plasma state by a pulse-laser beam for 60 min; 3) NiFe-LDHs nanosheets are formed by supercooling and subsequent condensation of the plasma.[35] Gao *et al.* expanded the application of laser ablation technique to synthesize CoO/CoFe-LDHs nanocomposites in sodium chloride solution.[36] Herein, during the irritation process, Co-Fe alloy target was activated and reacted with chloride and hydroxide ions to generate CoFe-LDHs nanosheets, and at the same time, excessive $Co^{2+}$ could interact with water to form CoO on the surface of CoFe-LDHs. It is believed that the CoFe-LDHs content in composites was determined by the concentration of sodium chloride. Except for the laser ablation method, Chen *et al.* first proposed a ultrasonication/mechanical stirring method to obtain NiCr-LDHs nanosheets without adding any growth inhibitor, in which the ultrasonication/mechanical stirring treatment acted as a mechanical inhibitor to greatly reduce the thickness of NiCr-LDHs during the co-precipitation process.[37] Afterwards, Wang *et al.* used the same method to successfully synthesize NiFe-LDHs nanosheets and studied the influence of ultrasonic time on their thickness.[38] With increasing ultrasonic time, the thickness of NiFe-LDHs decreased from 61.3 to 7.4 nm. However, excessive ultrasonic treatment partially destroyed the nanosheet structure of NiFe-LDHs, leading to fragmental morphology and increased thickness. Furthermore, Wang *et al.* applied a similar methodology to controllably grow NiCo-LDHs nanosheets on the surface of porous carbon, which were derived from *Dicranopteris dichotoma* stems.[39] As seen in Figure 5c, porous carbon and $Ni^{2+}$/$Co^{2+}$ ions were mixed in a beaker and the titration process was carried out under the ultrasonication/mechanical stirring. The mechanical treatment could effectively prevent the aggregation of NiCo-LDHs nanosheets to realize their rational distribution on porous carbon. Considering the simplicity and high efficiency, the above-mentioned mechanical strategies are promising for large-scale preparation of TM-LDHs nanosheets.

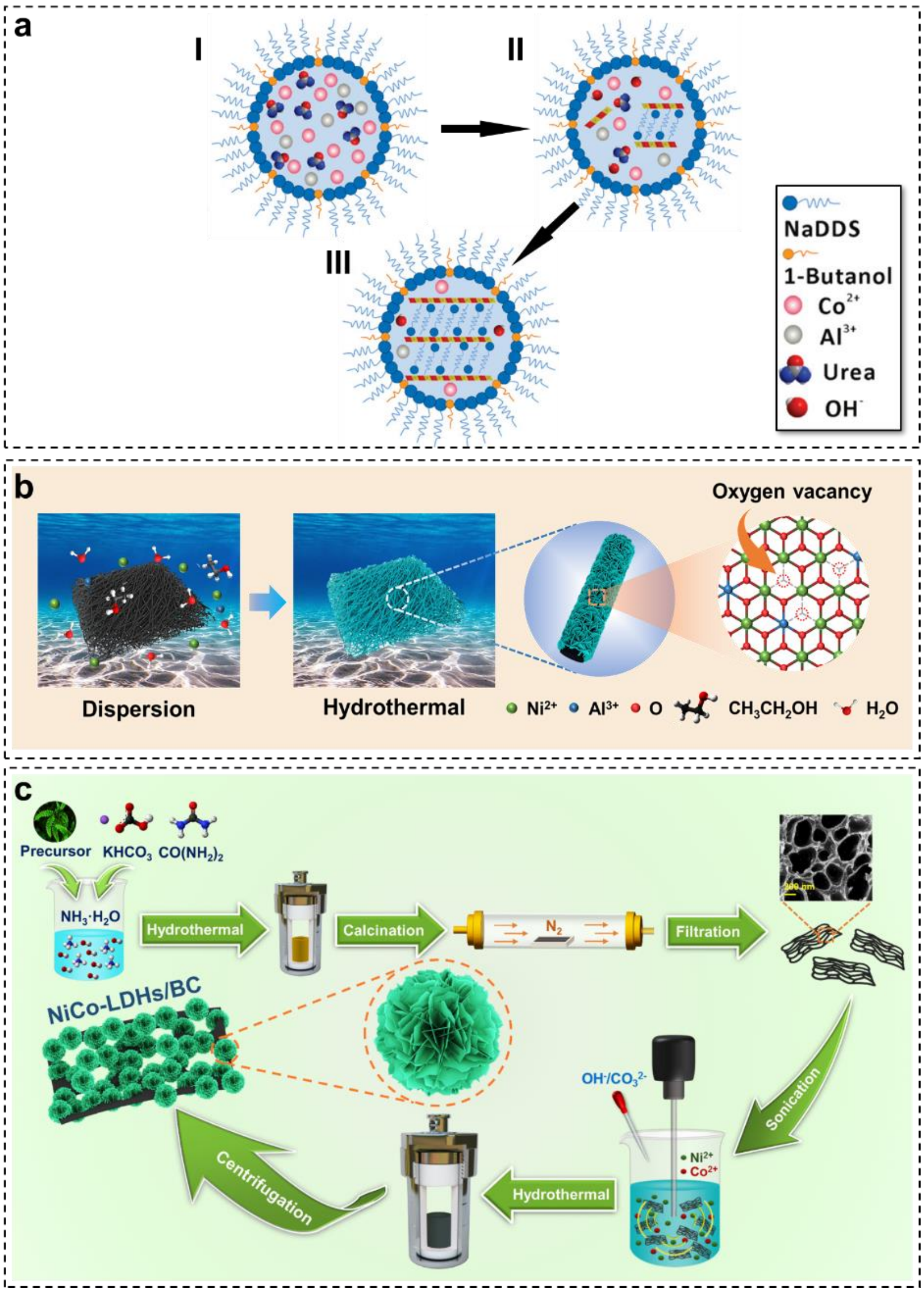

**Figure 5.** (a) Schematic representation of the formation of TM-LDHs nanosheets in microemulsion. Reproduced with permission.[27] Copyright 2011, American Chemical Society. (b) One-step hydrothermal strategy used for fabricating oxygen-deficient NiAl-LDH nanosheets. Reproduced with permission.[30] Copyright 2022, Science China Press. (c) *In situ* growth of NiCo-LDHs nanosheets on porous carbon materials. Reproduced with permission.[39] Copyright 2022, KeAi Publishing Communications Ltd.

In the following, the utilization of these fabricated TM-LDHs nanosheets for oxygen evolution reaction (OER), hydrogen evolution reaction (HER), urea oxidation reaction (UOR), nitrogen reduction reaction (NRR), small molecule oxidation, and biomass derivatives upgrading are articulated through systematically discussing the corresponding fundamental design principles and reaction mechanism.

# 3. Electrocatalytic Water Splitting

## 3.1. Oxygen Evolution Reaction

### *3.1.1. Mechanism*

In state-of-art electrochemical energy systems (e.g., fuel cell, water electrolyzer, metal-air battery), OER is the key reaction to determine their overall performance.[40] Commonly, OER with sluggish four-electron transfer proceeds through absorption/desorption of different oxygen intermediates. Deciphering the mechanism of OER is beneficial to develop efficient and robust OER electrocatalysts.

At present, two mainstream pathways of OER have been proposed as adsorbate evolution mechanism (AEM) and lattice oxygen mediated mechanism (LOM). Figure 6 distinctly illustrates these pathways. For the AEM pathway (Figure 6a), $OH^-$ undergo adsorption/deprotonation/coupling/desorption steps to release $O_2$ with the generation of OH*, O* and OOH* intermediates.[41] These four steps are all thermodynamically unfavorable, resulting in a high overpotential to drive OER.[42] Based on theoretical calculation and experimental results, the binding energies of OH*, O* and OOH* intermediates are found to be linearly correlated on the basis of the following relationship, $\Delta G_{OOH*} = \Delta G_{OH*} + 3.2 \pm 0.2$ eV.[43] Considering the overall energy barrier of deprotonation/coupling step (2.46 eV), the overpotential limit of OER is determined as $(3.2-2.46)/2 = 0.37$ V. Nevertheless, the overpotential of substantial reported OER electrocatalysts is lower than the limited value of AEM pathway. Under this circumstance, a LOM pathway to trigger activated lattice oxygen in OER electrocatalysts is proposed, and further classified into oxygen-vacancy, single-metal, and dual-metal site mechanisms (OVSM, SMSM, and DMSM).[44] For the OVSM pathway (Figure 6b), OOH* intermediates are initially formed on single metal active sites by combining activated lattice oxygen and $OH^-$.[45] After releasing oxygen molecules, the resulting oxygen vacancies are refilled by $OH^-$ to regenerate these active sites. Alternatively, the single metal active site in the SMSM pathway (Figure 6c) firstly absorbs $OH^-$ to generate OH* and O* intermediates following the first and second steps of AEM, respectively. Afterwards, O* intermediates are coupled with the activated lattice oxygen to form OO* intermediates, which are subsequently oxidized to release $O_2$.[46] Apart from single active site, $O_2$ in the DMSM pathway (Figure 6d) are generated from OO* intermediates, which are formed by direct coupling of adjacent lattice oxygen from two metal active sites.[47] Albeit LOM pathway involving O-O coupling bypass the theoretical limit of AEM pathway, the deprotonation energy is inevitably increased with the switch from AEM pathway to LOM pathway.[48] For this reason, new pathways with advantages of both AEM and LOM should be further explored.

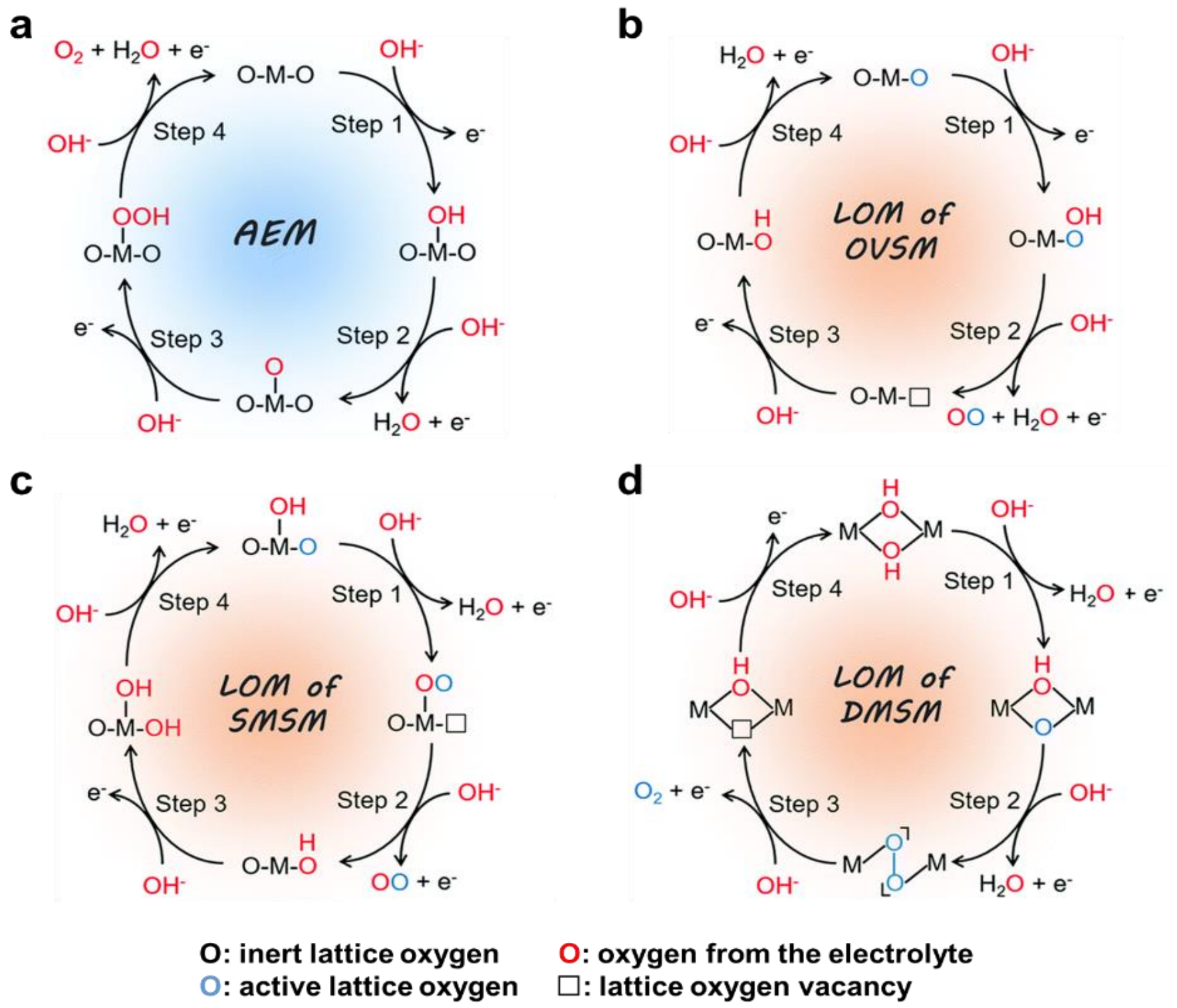

**Figure 6.** The schematic illustration of AEM and LOM pathways for alkaline OER. Reproduced with permission.[44] Copyright 2021, Royal Society of Chemistry.

*3.1.2. Research*

TM-LDHs nanosheets have been extensively investigated as efficient electrocatalysts for OER. At the early stage, pristine TM-LDHs nanosheets present outperformed OER activity in comparison with those commercial noble metal electrocatalysts. For example, Song *et al.* exfoliated a series of TM-LDHs (NiCo-, NiFe-, and CoCo-LDHs) in formamide to achieve nanosheets, which exhibited excellent OER activity.[10b] In comparison with bulk TM-LDHs, the enhanced electrocatalytic activity of nanosheets were mainly originated from the exposure of more active sites and the higher electronic. As such, NiFe-LDHs nanosheets achieved the best OER activity with an overpotential of 302 mV, which was 36 mV lower than that of $IrO_2$. The role of transition metal for OER was then unveiled by systematically studying four different transition metal (oxy)hydroxides.[49] Experimental results demonstrated that OER activities of transitional metal elements were negatively correlated to their OH-$M^{2+\delta}$ energetic strength. The trend established the fundamental criteria for designing active center of TM-LDHs nanosheets.

Later on, versatile modification strategies including surface engineering, strain engineering, doping, and vacancy introduction are employed to ameliorate the $O_2$ electrocatalytic evolution ability of TM-LDHs nanosheets. Owing to the nearly-identical length of Ni-O and Fe-O bonds, the synergistic interaction between Ni and Fe sites, and high electronic conductivity,[50] it is noticed that recent reported TM-LDHs nanosheets for OER are mostly constructed on the basis of NiFe-LDHs. For example, Zhao *et al*. found that (012) edge plane possessed the lowest onset overpotential for OER through theoretical calculation methods (Figure 7a).[51] According to the theoretical results, NiFe-LDHs nanosheet-arrays-on-microplates (NiFe NSAs-MPs) with abundant exposed (012) crystal planes were fabricated and the corresponding scanning electron microscopy (SEM) image is shown in Figure 7b. As expected from linear sweep voltammetry (LSV) curves in Figure 7c, the overpotential of NiFe NSAs-MPs at 100 mA $cm^{-2}$ was only 250 mV, which was far below the overpotential of (003) plane-dominated NiFe-LDHs microsheet arrays (NiFe-LDHs-MSAs) and pure nickel foam (NF). Afterwards, abundant dislocations are validated to promote the OER active phase formation of NiFe-LDHs nanosheets.[52] Compare with pristine NiFe-LDHs nanosheets, the overpotential of NiFe-LDHs nanosheets with ample dislocations at 10 mA $cm^{-2}$ was reduced from 223 to 199 mV, manifesting the positively influence of dislocation on OER performance.

Doping and vacancy generation are common methods to tune the electronic structure of TM-LDHs nanosheets at the atomic or molecular level, and thus modulate their inherent OER activity. For example, Zeng *et al*. successfully incorporated Ru atoms onto the surface of CoV-LDHs nanosheets to enhance OER activity.[53] According to X-ray absorption near-edge structure (XANE) results (Figure 7d), the addition of $Ru^{3+}$ reduced the valence state of vanadium element to cause the local charge redistribution, which strengthened the electronic coupling between Ru atoms and CoV-LDHs nanosheets. As a result, the CoVRu-LDHs nanosheets only required an overpotential of 296 mV to gain a current density of 50 mA $cm^{-2}$, while CoV-LDHs and $RuO_2$ required 395 and 353 mV, respectively (Figure 7e). Meanwhile, according to the d-band center positions of CoV-LDHs and CoVRu-LDHs in Figure 7f, the upshift implied the existence of more unoccupied states, which was prone to capture electrons. Both experimental and theoretical results confirmed the beneficial effect of Ru doping on OER performance. Moreover, Xie *et al*. clarified the role of metal vacancies on OER activity.[54] Concretely, after comparing free energy plots of NiFe-LDHs nanosheets with M(II) and M(III) defects, the overpotential of Ni(II)-Fe(III) sites was 0.2 eV lower than that of Ni(II)-Ni(II) sites, denoting that M(II) defects were more effective during OER process.

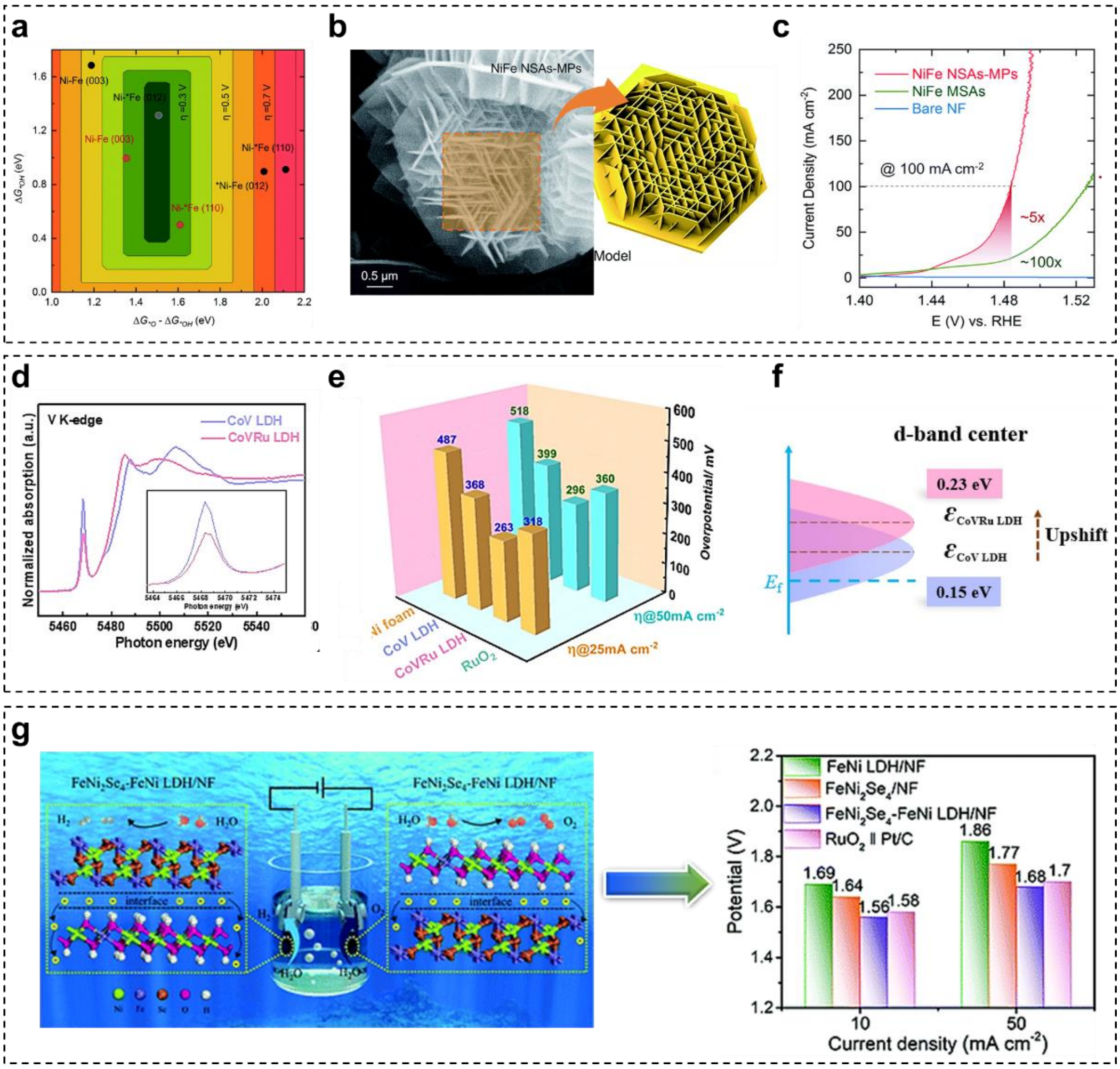

**Figure 7.** (a) Theoretical overpotential plot for different crystal facets of NiFe-LDHs. (b) SEM image and corresponding schematic model of NiFe NSAs-MPs. (c) LSV curves of NiFe NSAs-MPs, NiFe MSAs and NF in 1 M KOH electrolyte at 5 mV $s^{-1}$. Reproduced with permission.[51] Copyright 2021, Royal Society of Chemistry. (d) V K-edge normalized XANES spectra of CoV-LDHs and CoVRu-LDHs nanosheets. (e) Overpotential values of CoV-LDHs, CoVRu-LDHs, $RuO_2$, and NF at 25 and 50 mA $cm^{-2}$. (f) d-band center of CoV-LDHs and CoVRu-LDHs. Reproduced with permission.[53] Copyright 2023, Elsevier. (g) Cell voltage for a symmetrical

$FeNi_2Se_4$-FeNi-LDHs based electrolyzer at current densities of 10 and 50 mA cm$^{-2}$. Reproduced with permission.[55] Copyright 2022, Royal Society of Chemistry.

To accelerate the kinetic rate of OER, TM-LDHs nanosheets are always coupled with other materials, such as carbon materials. It is found that Co-C@NiFe-LDHs nanofibrous electrocatalyst exhibited better electrocatalytic activity than and expensive $RuO_2$, which was ascribed to the fast charge transfer between NiFe-LDH nanosheets and Co-C.[56] Furthermore, heterogeneous interface engineering on TM-LDHs nanosheets has been also broadly explored to improve their OER activity. For example, Ouyang *et al*. combined *p*-type CuO nanowires with n-type FeCoNi-LDHs nanosheets to construct heterostructured electrocatalyst for OER.[57] Owing to the difference of Fermi levels between FeCoNi-LDHs and CuO, an electric field was generated at the heterogeneous interface to induce charge transfer within the whole electrocatalyst and thereby promote the adsorption ability of $OH^-$, giving rise to elevated OER activity. Yu *et al*. fabricated $FeNi_2Se_4$-FeNi-LDHs bifunctional electrocatalyst for water splitting via partial selenization of FeNi LDHs nanosheets.[55] Similarly, coupling counterparts in $FeNi_2Se_4$-FeNi-LDHs optimized the electronic environment around active center to enhance their electrocatalytic activities. Therefore, a symmetrical $FeNi_2Se_4$-FeNi-LDHs based electrolyzer could deliver 10 mA cm$^{-2}$ by applying a cell voltage of 1.56 V (Figure 7g).

**Table 1.** Recent reported TM-LDHs nanosheets for OER.

| Materials | Electrolyte | η (mV @ mA cm$^{-2}$) | Tafel slope (mV dec$^{-1}$) | Refs. |
|---|---|---|---|---|
| CoV-LDHs | 1 M KOH | 250 @ 10 | 44 | [58] |
| NiFe-LDHs | 1 M KOH | 270 @ 10 | 48.6 | [59] |
| NiFe-LDHs-MPs | 1 M KOH | 250 @ 100 | 34.5 | [51] |
| NiFe-LDHs | 1 M KOH | 324 @ 10 | 57.4 | [60] |
| CoNi-LDHs-E | 1 M KOH | 280 @ 10 | 81 | [61] |
| Ta-NiFe LDHs | 1 M KOH | 260 @ 50 | 58.95 | [62] |
| D-NiFe LDHs | 1 M KOH | 199 @ 10 | 26.9 | [52] |
| NiFe LDHs/$Co_{1-x}S$ | 1 M KOH | 251 @ 10 | 41.67 | [63] |
| NiFeV-LDHs | 1 M KOH | 195 @ 20 | 42 | [64] |
| $NiFe_{III}$(1:1)-LDHs | 1 M KOH | 183 @ 10 | 31.1 | [65] |
| $NiCo_1Fe_1$-LDHs | 1 M KOH | 231 @ 10 | 59 | [66] |
| CoVRu-LDHs | 1 M KOH | 263 @ 25 | 74.5 | [53] |
| Fe-NiV-LDHs | 1 M KOH | 255 @ 10 | 56 | [67] |
| NiFeV-LDHs | 1 M KOH | 224 @ 10 | 32.7 | [68] |
| NiFeNb-0.25-LDHs | 1 M KOH | 277 @ 100 | 50.6 | [69] |
| NiFeCo-LDHs | 1 M KOH | 249 @ 10 | 42 | [70] |
| v-NiFe-LDHs | 1 M KOH | 195 @ 10 | 47.9 | [71] |
| EE-NiFe-LDHs | 1 M KOH | 205 @ 10 | 41.8 | [72] |
| NiCoFe-LDHs | 1 M KOH | 174 @ 10 | 50 | [73] |
| NiFe-LDHs-$V_{Ni}$ | 1 M KOH | 229 @ 10 | 62.9 | [74] |
| v-NiFe-LDHs | 1 M KOH | 150 @ 10 | 37.1 | [75] |
| D-NiFeZn-LDHs | 0.1 M KOH | 200 @ 20 | 34.9 | [54] |
| NiFe-LDHS-$V_O$ | 1 M KOH | 230 @ 10 | 39.6 | [76] |
| D-CoFe-LDHs | 1 M KOH | 283 @ 10 | 39 | [77] |
| MnNiFe-LDHs-laser | 1 M KOH | 220 @ 10 | 37 | [78] |
| M-NiFe-LDHs | 1 M KOH | 217 @ 10 | 45.1 | [79] |
| AGC/MnCo-LDHs | 1 M KOH | 370 @ 10 | 127.5 | [80] |
| Co-C@NiFe-LDHs | 1 M KOH | 249 @ 10 | 57.9 | [56] |
| FeNi-LDHs/CoP | 1 M KOH | 231@ 20 | 33.5 | [81] |
| FeCoNi-LDHs/CuO/Cu | 1 M KOH | 243.1 @ 50 | 63.8 | [57] |
| $FeNi_2Se_4$-FeNi-LDHs | 1 M KOH | 205 @ 10 | 30.14 | [55] |
| MIM-CoFe-LDHs | 1 M KOH | 216.8 @ 10 | 39.3 | [82] |
| CoNi-LDHs/$Ti_3C_2T_x$ | 1 M KOH | 200 @ 50 | 68 | [83] |
| CoNi-LDHs@PCPs | 1 M KOH | 350 @ 10 | 58 | [84] |
| $Ni_3S_2$/Cu-NiCo-LDHs | 1 M KOH | 119 @ 10 | 70 | [85] |
| NiCo-LDHs/NiCoS | 1 M KOH | 308@ 100 | 48 | [86] |
| A-NiFe-LDHs/CuS | 1 M KOH | 240 @ 100 | 41 | [87] |
| NiCo@NiFe-LDHs | 1 M KOH | 184 @ 10 | 46.1 | [88] |
| $Fe_2O_3$/NiFe-LDHs | 1 M KOH | 220 @ 500 | 32 | [9b] |
| NiFe-LDH/V-$Co_4N$ | 1 M KOH | 203 @ 10 | 26 | [89] |
| Mo-$NiS_x$@NiFe-LDHs | 1 M KOH | 271 @ 100 | 44.41 | [90] |
| NF-$Ni_2P$@NiFe LDHs | 1 M KOH | 203.5 @ 10 | 48.29 | [91] |
| $NiCo_2S_4$@NiFe-LDHs | 1 M KOH | 287 @ 10 | 86.4 | [92] |
| $Ni_3S_2$/VG@NiCo-LDHs | 1 M KOH | 350 @ 100 | 65 | [93] |

According to Table 1, all TM-LDHs nanosheets exhibit high OER activity in alkaline electrolyte. At present, the OER mechanism of most TM-LDHs nanosheets follow AEM pathway, while LOM pathway is appropriate for TM-LDHs nanosheets with modulated lattice oxygen coordination environment.[79,94] In a practical OER process, the low intrinsic diffusivity of activated lattice oxygen from bulk to surface could lead to the alternative mechanistic transformation between LOM and AEM, resulting in insufficient utilization of active sites.[48] Thus the difference in OER performance between AEM- and LOM-based TM-LDHs nanosheets is indistinct. With this consideration, the morphology of OER electrocatalysts should be designed to optimize their electronic/ionic transfer ability. Moreover, in order to further enhance the intrinsic OER activity of TM-LDHs nanosheets, novel mechanism is still needed to completely break the scaling relationship between OH* and OOH* intermediates.

## 3.2. Hydrogen Evolution Reaction

### *3.2.1. Mechanism*

Hydrogen has been considered as an ideal fuel because of its extremely high heating value (142 MJ $kg^{-1}$), earth abundance, and environmental benignity.[95] In natural environment, water and hydrocarbon are two primary sources for hydrogen. Currently, industrial hydrogen production mainly relies on steam reforming of hydrocarbons, which consumes fossil fuels and release plenty greenhouse gas.[96] In contrast, water electrolysis with zero emission and high hydrogen purity is regard as a more promising approach.[97]

HER is the cathodic reaction of water electrolysis, which involves two-electron transfer. Since HER performance is pertinent on the pH value of electrolyte, the mechanism of HER needs to be elucidated in different electrolytes (Figure 8).[98] In acidic medium, a proton is initially adsorbed on active sites to form H* (Volmer step). Subsequently, H* is combined with either another H* (Tafel step) or another proton (Heyrovsky step) to release hydrogen. Generally, the Tafel slope value is a classical indicator to determine rate-demining step (RDS) in HER.[99] For Volmer, Tafel, and Heyrosky steps, the corresponding Tafel slope values are estimated as 120, 30, and 40 mV $dec^{-1}$ [100]. According to these values, the mechanism of HER can be separated into Volmer-Tafel and Volmer-Heyrosky mechanism. The reaction pathway in neutral or alkaline medium is nearly identical to that in acidic medium except that water molecules are participated to provide protons. However, because of the additional water dissociation step, the HER process in neutral and alkaline electrolytes is reported as 2−3 order of magnitude slower than that in acidic electrolyte.[101] Consequently, both water dissociation and hydrogen adsorption should be considered in designing efficient electrocatalysts towards neutral and alkaline HER.

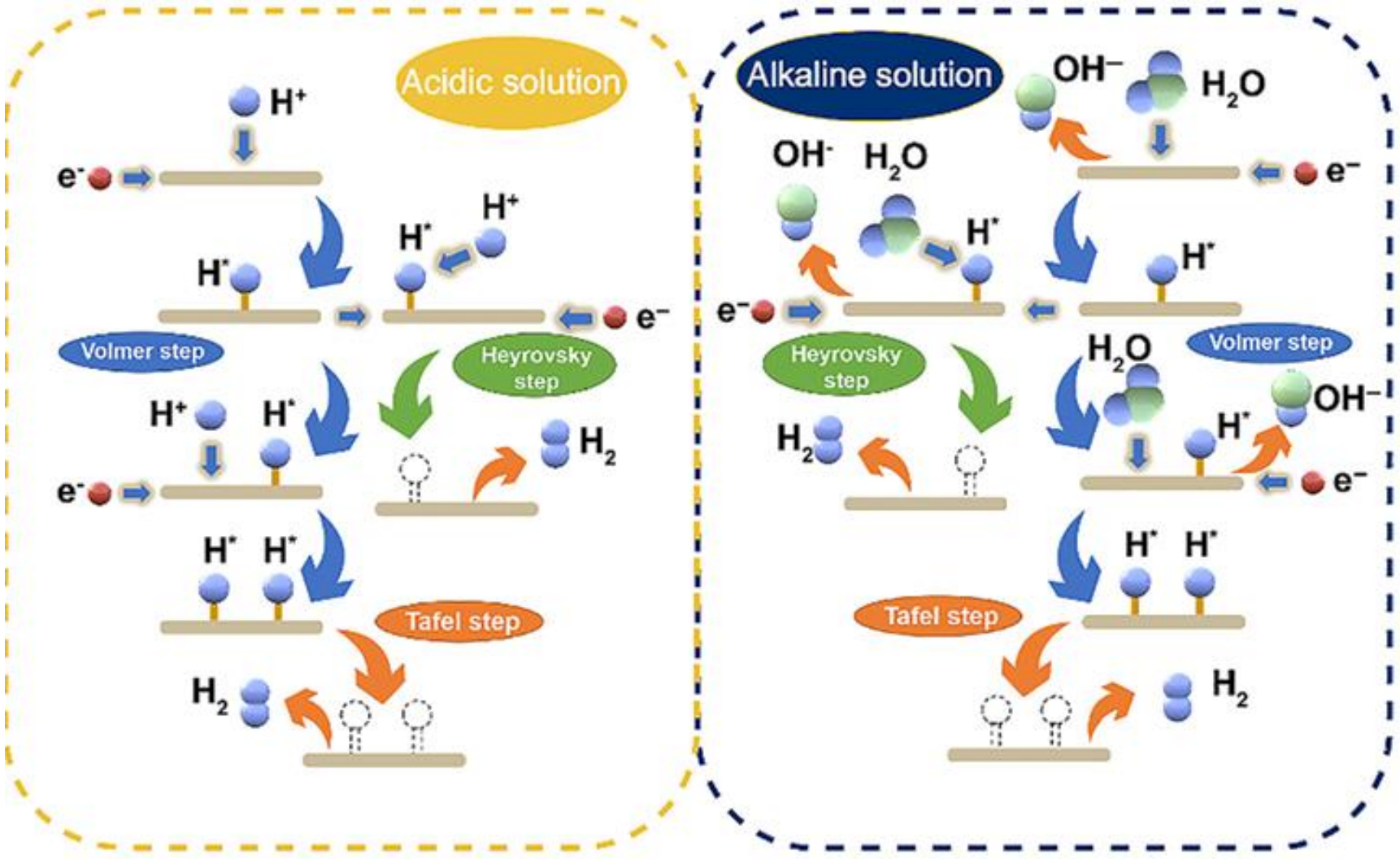

**Figure 8.** The schematic illustration of HER mechanism in acidic and alkaline media. Reproduced with permission.[98b] Copyright 2022, Wiley.

*3.2.2 Research*

Over the past few decades, remarkable endeavors have been devoted to construct non-precious TM-LDHs nanosheets to replace noble metal-based electrocatalysts. However, HER performance of pristine TM-LDHs nanosheets is not desirable in alkaline electrolyte owing to the sluggish kinetics of Volmer step. Thus, various strategies have been utilized to improve their HER performance.

HER activity of TM-LDHs nanosheets could be enhanced through defect engineering, such as doping metal elements and constructing oxygen vacancies. Chen *et al*. doped noble metal Ru into NiFe-LDHs (NiFeRu-LDHs) nanosheets to achieve a superior HER performance.[102] As shown in Figure 9a, overpotential of NiFeRu-LDHs nanosheets for HER was only 29 mV to drive a current density of 10 mA $cm^{-2}$. Afterwards, the superior HER activity was elucidated by density functional theory (DFT) calculations (Figure 9b and Figure 9c). After Ru doping, energy barrier of the Volmer step was decreased from 1.02 to 0.5 eV, suggesting that the HER kinetic process was greatly accelerated. The beneficial effect of Ru doping was also validated on NiV-LDHs nanosheets.[103] Theoretical calculations suggested that Ru doping could significantly decrease the hydrogen adsorption energy in both Volmer and Heyrovsky steps, thus boosting the HER kinetics. Furthermore, Liu *et al*. synthesized Ce doped CoFe-LDHs nanosheets with abundant oxygen vacancies (V-Ce/CoFe-LDHs) via a plasma etching strategy.[104] Herein, Ce doping and oxygen vacancy not only lowered the energy barrier of HER, but also led to more active sites and improved electrical conductivity, thus giving rise to an excellent HER performance.

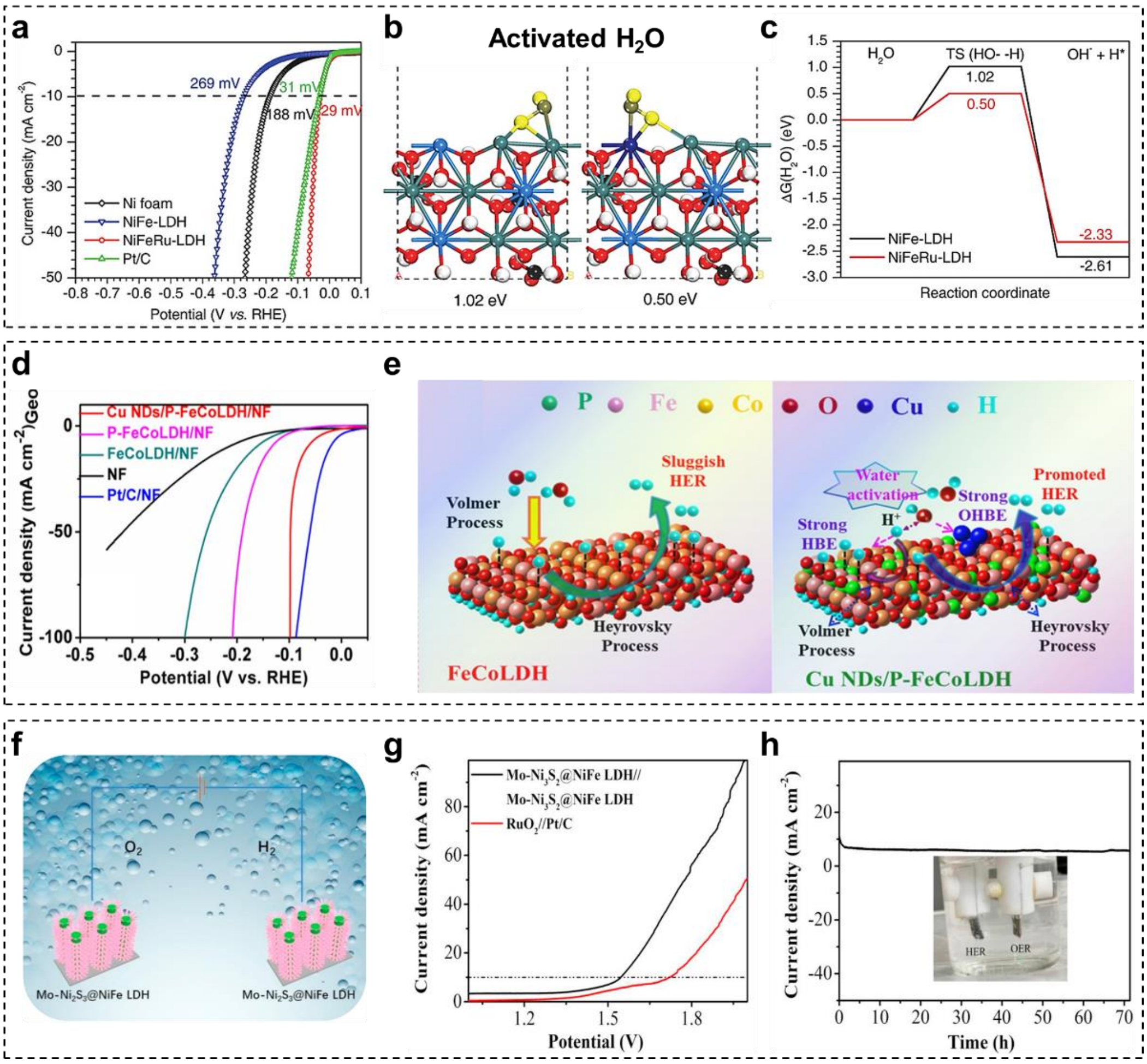

**Figure 9.** (a) LSV curves of NiFeRu-LDHs, NiFe-LDHs, NF, and Pt/C electrocatalysts for HER. (b) Adsorption free energies of activated $H_2O$, and (c) corresponding free energy diagrams of NiFe-LDHs and NiFeRu-LDHs nanosheets. Reproduced with permission.[102] Copyright 2018, Wiley. (d) LSV curves of Cu NDs/P-FeCo-LDHs/NF, P-FeCo-LDHs/NF, FeCo-LDHs/NF, Pt/C/NF, and NF with a scan rate of 5 mV $s^{-1}$ for HER. (e) Schematic representation of alkaline HER mechanism on FeCo-LDHs and Cu NDs/P-FeCo-LDHs. Reproduced with permission.[105] Copyright 2021, Elsevier. (f) Schematic of Mo-$Ni_2S_3$@NiFe-LDHs based symmetrical electrolyzer for water splitting. (i) LSV curves of the symmetrical electrolyzer and $RuO_2$//Pt/C couple. (j) Stability test of the symmetrical electrolyzer over 72 h. Reproduced with permission.[106] Copyright 2021, Elsevier.

Hybridization of TM-LDH materials with highly conductive materials is another strategy to expedite HER kinetics. Taking graphene oxide (GO) as example, the appropriate quantity of GO can control the growth of NiFe-LDHs nanosheets during the electrodeposition process.[107] The as-synthesized composites possessed a large electrochemical surface area, strong electronic interaction between NiFe-LDHs and GO, which are advantageous for hydrogen generation. In addition, Feng *et al*. employed P doping and subsequent Cu nanodots (NDs) deposition to delicately construct efficient Cu NDs/P-FeCo-LDHs electrocatalyst for HER.[105] The Cu NDs/P-FeCoLDHs exhibited Pt-like HER activity with an overpotential of 63 mV at 10 mA $cm^{-2}$ (Figure 9d), which is much lower than those of their counterparts. According to mechanistic analysis, P doping and deposited Cu NDs were responsible for optimizing H* binding energy and accelerating water dissociation on the surface of electrocatalyst, respectively. Thus, the combined method can effectively expedite the kinetics of alkaline HER (Figure 9e).

Furthermore, TM-LDHs based hybrid electrocatalysts have been used in a practical electrolyzer with high $H_2$ production efficiency. As shown in the Figure 9f, heterostructured Mo-$Ni_2S_3$@NiFe-LDHs electrocatalyst was utilized in a symmetrical alkaline electrocatalytic cell for water dissociation.[106] Surprisingly, in comparison with Pt/C//$RuO_2$, the Mo-$Ni_2S_3$@NiFe-LDHs based electrolyzer only needed 1.54 V to accomplish 10 mA $cm^{-2}$ (Figure 9g). Also, the symmetric electrolyzer exhibited fabulous durability for 72 h (Figure 9h).

**Table 2.** Recent reported TM-LDHs nanosheets for HER.

| Materials | Electrolyte | η (mV @ mA $cm^{-2}$) | Tafel slope (mV $dec^{-1}$) | Refs. |
|---|---|---|---|---|
| NiFeRu-LDHs | 1 M KOH | 29 @ 10 | 31 | [102] |
| NiMoCo-LDHs | 1 M KOH | 123 @ 10 | 67.39 | [108] |
| NiVRu-LDHs | 1 M KOH | 12 @ 10 | 40 | [103] |
| NiFeAu-LDHs | 1 M KOH | 89 @ 10 | 90 | [109] |
| La-NiFe LDHs-3 | 1 M KOH | 57 @ 10 | 73.7 | [110] |
| Ir–NiCo LDHs | 1 M KOH | 21 @ 10 | 33.2 | [111] |
| Pt@CuFe-LDHs | 1 M KOH | 33 @ 10 | 34 | [112] |
| V-Ce/CoFe-LDHs | 1 M KOH | 73 @ 10 | 69 | [104] |
| A-NiCo-LDHs/NF | 1 M KOH | 151 @ 100 | 57 | [113] |
| CoFe-LDHs-F | 1 M KOH | 166 @ 10 | 92 | [114] |
| $Co_3Fe_1$-LDHs/rGO | 1 M KOH | 110 @ 10 | 106 | [115] |
| GO-NiFe-LDHs | 1 M KOH | 119 @ 10 | 36 | [107] |
| H-$CoS_x$@NiFe LDHs | 1 M KOH | 95 @ 10 | 90 | [116] |
| $Ni_3Se_4$@NiFe LDHs | 1 M KOH | 85 @ 10 | 98.6 | [117] |
| Cu@NiFe-LDHs | 1 M KOH | 116 @ 10 | 58.9 | [118] |
| Ag@NiFe-LDHs | 1 M KOH | 78 @ 10 | 58.1 | [119] |
| Cu NDs/P-FeCo-LDHs | 1 M KOH | 63 @ 10 | 41.74 | [105] |
| Mo-$Ni_2S_3$@NiFe LDHs | 1 M KOH | 109 @ 10 | 36.5 | [106] |

As summarized in Table 2, most TM-LDH nanosheets present high HER efficiency in alkaline electrolyte. According to their Tafel slope values, the desorption of hydrogen gas in Tafel and Heyrosky steps are determined as main RDS for alkaline HER. Thus the modulation of H* desorption energy is crucial for designing TM-LDHs nanosheets as high-performance HER electrocatalysts. Besides, long-term stability of TM-LDHs nanosheets at large current density is still far from satisfaction. Combining the state-of-art fabrication methods, the construction of TM-LDHs nanosheet based hybrid structure with optimal hydrogen adsorption/desorption ability is expected to realize long-term efficient hydrogen production.

## 4. Urea Oxidation/Nitrogen Reduction Reaction

### 4.1. Urea Oxidation Reaction

#### *4.1.1. Mechanism*

Urea electrolysis ($CO(NH_2)_2 + H_2O \rightarrow N_2 + 3H_2 + CO_2$) is promising to replace conventional water electrolysis for hydrogen production due to its low theoretical electrocatalytic voltage (0.37 *vs.* 1.23 V) and abundant natural sources.[120] UOR is an anodic reaction with sluggish six-electron transfer, which determines the overall electrocatalytic performance.[121] Generally, UOR is favorable to proceed in alkaline electrolyte since neutral and acid electrolytes can accelerate the corrosion of metallic electrocatalysts.[122]

For the alkaline UOR, nickel-based electrocatalysts have been mostly reported due to their low-cost and intriguing UOR performance. Initially, NiOOH was considered as active species of nickel-based electrocatalysts and two possible mechanisms were proposed : 1) direct oxidation mechanism $[NiOOH{\cdot}CO(NH_2)_2]_{ads} + 6OH^- \rightarrow [NiOOH{\cdot}CO_2]_{ads} + N_2 + 5H_2O + 6e^-$; 2) indirect oxidation mechanism $6NiOOH + CO(NH_2)_2 + H_2O \rightarrow 6Ni(OH)_2 + N_2 + CO_2$ (Figure 10a).[123] Afterwards, the assumptions were investigated by *in situ* tools. Vedharathinam *et al.* utilized *in situ* time resolved Raman spectroscopy to study the phase transformation of NiOOH during the alkaline UOR process under open-circuit conditions.[124] In pure alkaline electrolyte, two characteristic peaks of NiOOH at 479/559 $cm^{-1}$ remained unchanged for 13 min. With the addition of urea, a new peak of $Ni(OH)_2$ at 528 $cm^{-1}$ was detected after 4 min. These comparative results indicated that NiOOH could be reduced to $Ni(OH)_2$ during alkaline UOR, which was also verified by *in situ* X-ray diffraction technique.[125] Therefore, the mechanism of alkaline UOR is confirmed as indirect catalyst regeneration mechanism. To better understand alkaline UOR, Daramola *et al*. calculated three possible reaction pathways of urea dissociation with NiOOH using DFT calculations.[126] After analyzing rate constants and free energies of intermediate steps, the optimized pathway was proposed as $M{\cdot}CO(NH_2)_2 \rightarrow M{\cdot}CO(NH{\cdot}NH_2) \rightarrow M{\cdot}CO(NH{\cdot}NH) \rightarrow M{\cdot}CO(NH{\cdot}N) \rightarrow M{\cdot}CO(N_2) \rightarrow M{\cdot}CO(OH) \rightarrow M{\cdot}CO(OH{\cdot}OH) \rightarrow M{\cdot}COO$ (M = NiOOH) and the last step of $CO_2$ desorption was found to be the RDS.

Recently, Chen *et al*. rebuilt the conventional indirect mechanism of UOR based on detailed experimental and theoretical results.[127] As seen in Figure 10b, β-$Ni(OH)_2$ is electrocatalytic dehydrogenated to generate β-Ni(OH)O intermediates, which could further react with urea via a spontaneous dehydrogenative oxidation process. For the second spontaneous reaction, β-Ni(OH)O intermediates initially seize the protons from two amine groups in one urea molecule and CO(NN) intermediates with an intramolecular N-N bond are formed through proton-coupled electron transfer (PCET). After complicated hydration, rearrangement, and PCET processes, CO(NN) intermediates are transformed into $N_2$ and $CO_2$. In sum, current understanding of UOR is limited and great efforts should be devoted to clarifying the UOR mechanism by employing advanced *in situ* characterization instruments and powerful theoretical calculations.

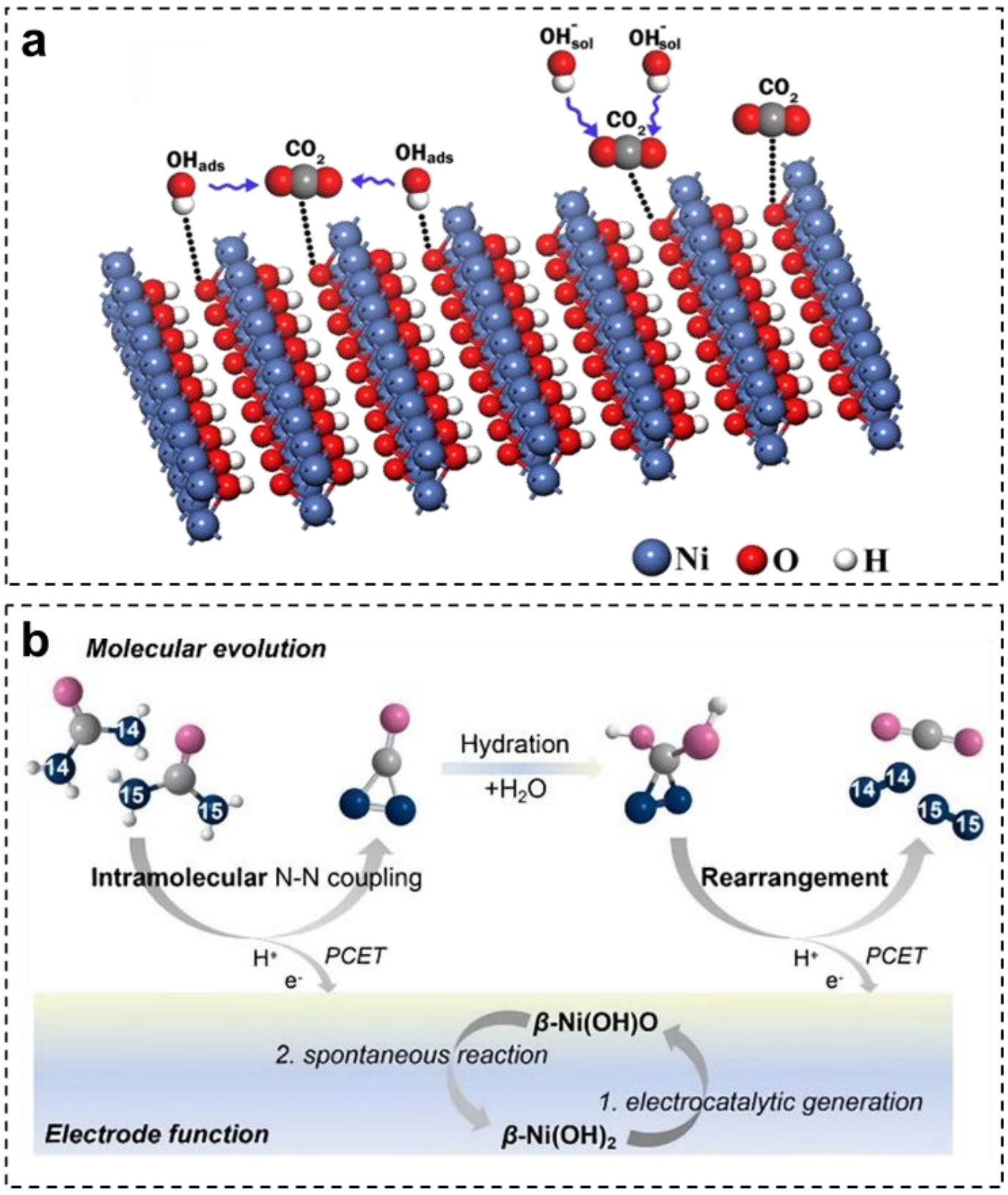

**Figure 10.** (a) Schematic illustration of indirect UOR mechanism. Reproduced with permission.[123a] Copyright 2016, Elsevier. (b) The representation of intramolecular N-N coupling mechanism for UOR. Reproduced with permission.[127] Copyright 2021, Wiley.

*4.1.2 Research*

TM-LDHs nanosheets have intensively emerged as electrocatalysts for UOR. Specifically, Zeng *et al*. employed a scalable liquid exfoliation method to obtain nitrate-exchanged NiCo-LDHs nanosheets for efficient UOR.[128] The synthetic method with long-term anion exchange process generated NiCo-LDHs nanosheets with well-defined hexagonal shape (Figure 11a). In comparison with carbonate-exchanged NiCo-LDHs and NiCo hydroxide, nitrate-exchanged NiCo-LDHs nanosheets displayed better UOR performance due to larger interlayer spacing. Moreover, the current efficiencies of OER and UOR were estimated via voltage step analyses and rotating ring-disk electrode (RRDE) tests. In the voltage step analyses, the cell voltage range was applied between NiCo-LDHs nanosheets and Pt foil from 1.35 to 1.60 V. As seen in Figure 11b, the current efficiencies of UOR were > 54% at all cell voltage values and achieved a maximal value of 88% at 1.50 V. In RRDE tests, a ring current of 7.6 μA was detected under the constant disk current of 200 μA in alkaline electrolyte with urea (Figure 11c). The corresponding current efficiency of UOR was calculated as around 80%, which is consistent with voltage step analyses. Wang *et al.* successfully applied a facile hydrothermal method to grow Cr doped CoFe-LDHs (CoFeCr-LDHs) nanosheets on NF substrate, which possessed an ultralow potential of 1.305 V *vs.* RHE for UOR at a current density of 10 mA $cm^{-2}$.[129] The outstanding UOR performance was considered to be originated from the inherent electron-withdrawing ability of Cr ions. Besides, a full electrolyzer was assembled with CoFeCr-LDHs nanosheets anode and Pt-C cathode to generate hydrogen. With the aid of urea, the cell voltage of the electrolyzer to achieve a current density of 10 mA $cm^{-2}$ was decreased from 1.467 to 1.329 V. Thus, efficient hydrogen production could be realized by driving such a device with a commercial 1.5 V battery.

Based on above experimental results, doping high-valence metal elements into TM-LDHs nanosheets is validated as an effective strategy to obtain superior UOR performance. Meanwhile, the beneficial effect of doping is also studied by many researchers in theoretical perspective. For example, Wang *et al.* designed Mo and V co-doped Ni-based LDHs (NiMoV-LDHs) nanosheets as highly efficient electrocatalysts for UOR and investigated the effect of dopants through DFT calculations.[130] Based on the total density of states (DOS) curves in Figure 11d, NiMoV-LDHs possessed the strongest DOS signal around the Fermi level, denoting that Mo and V-doping could significantly promote the electron transfer during UOR. Therefore, a urea electrolytic cell based on NiMoV-LDHs nanosheets showed superior UOR performance in various simulated urea-polluted electrolytes (Figure 11e). Except for doping, incorporating highly conductive carbon materials with TM-LDHs nanosheets is another strategy to improve UOR performance. For instance, Feng *et al.* constructed a novel hybrid UOR electrocatalyst by anchoring fullerene quantum dot decorated CoNi-LDHs nanosheets on porous NF substrate (FQD/CoNi-LDHs/NF), which required 1.42 V *vs.* RHE to deliver a current density of 100 mA $cm^{-2}$ for UOR.[131] Subsequently, the vital role of FQD was revealed by performing DFT calculations on CoNi-LDHs and FQD/CoNi-LDHs. After comparing the DOS profiles of these electrocatalysts, it is clearly concluded that FQD could increase the electronic conductivity and charge density, resulting in fascinating UOR performance of FQD/CoNi-LDHs/NF.

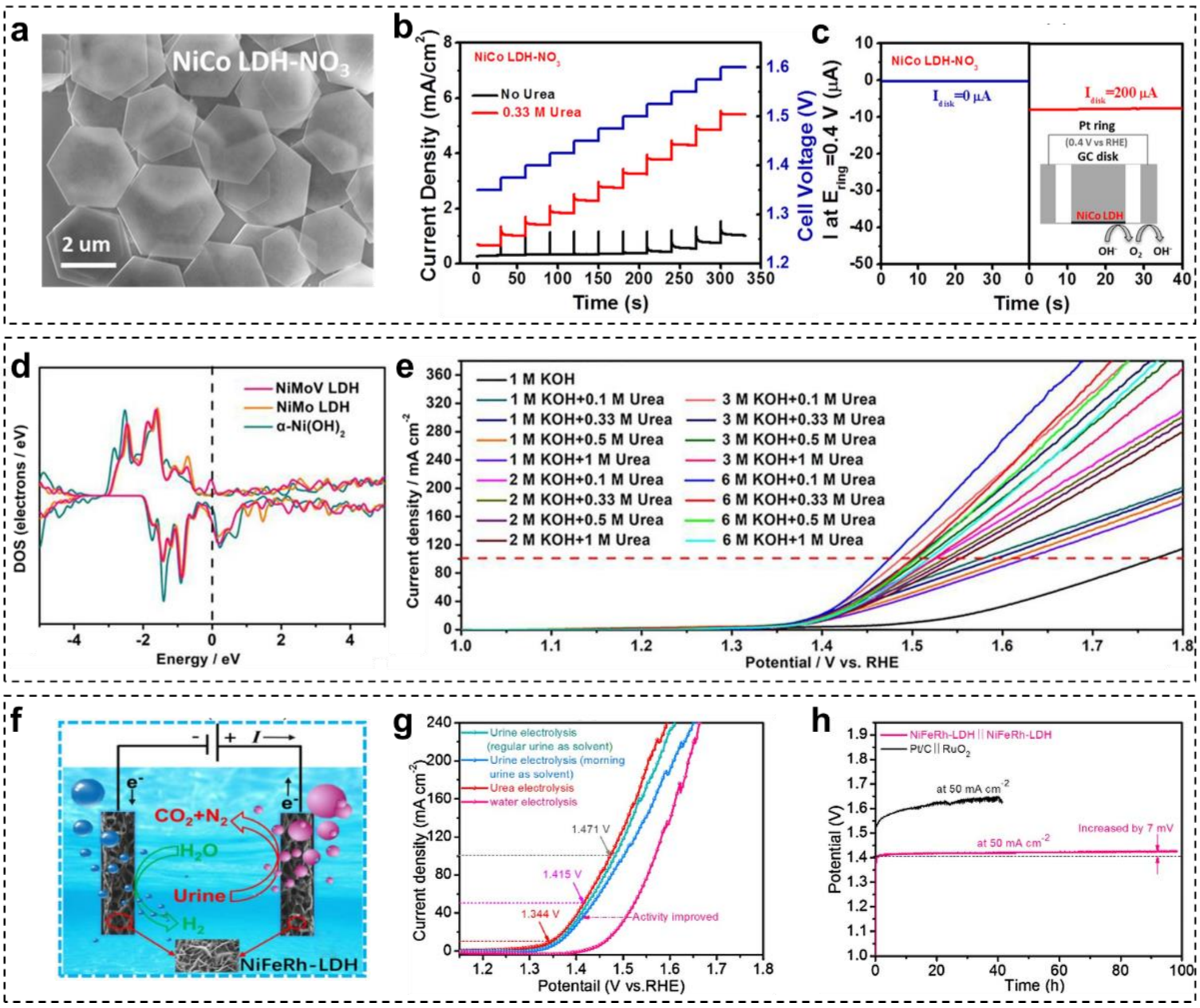

**Figure 11.** (a) SEM images of nitrate-exchanged NiCo-LDHs nanosheets. (b) Voltage step analyses in 1 M KOH with and without 0.33 M urea. (c) The ring current on RRDE in 1 M KOH + 0.33 M urea electrolyte with the ring potential of 0.4 V *vs.* RHE. Reproduced with permission.[128] Copyright 2019, American Chemical Society. (d) Calculated total DOS of NiMoV-LDHs/NF, NiMo-LDHs/NF and α-$Ni(OH)_2$/NF. (e) Polarization curves of a NiMoV-LDHs/NF based urea electrolytic cell in different electrolytes. Reproduced with permission.[130] Copyright 2022, Elsevier. (f) Schematic diagram of a symmetrical urea electrolytic cell using NiFeRh-LDHs nanosheets. (g) Comparison of CV curves for water, urea, and urine electrolysis. (h) Stability measurement of NiFeRh-LDHs nanosheets for urea electrolysis at 50 mA $cm^{-2}$. Reproduced with permission.[132] Copyright 2021, Elsevier.

To further extend the application of TM-LDHs nanosheets in UOR, Sun *et al.* fabricated bifunctional Rh-doped NiFe-LDHs (NiFeRh-LDHs) nanosheets via an ethylene glycol assisted hydrothermal method.[132] The introduction of Rh-doping optimized adsorption energy of H* and urea, which boosted HER and UOR performance of NiFeRh-LDHs nanosheets. Thus, NiFeRh-LDHs nanosheets could be applied as high-performance anode and cathode simultaneously in a symmetrical urea electrolytic cell (Figure 11f). Figure 11g presents the comparison of cyclic voltammetry (CV) curves for water, urea, and urine electrolysis using this novel designed cell. It is obviously seen that the electrolytic performance of urea and urine surpassed that of water significantly. Specifically, the cell voltages for urea electrolysis were required as 1.344 and 1.471 V to afford current densities of 10 and 100 mA cm$^{-2}$, respectively. Furthermore, the symmetric electrolytic cell could operate steadily at 50 mA cm$^{-2}$ for at least 100 h (Figure 11h), demonstrating excellent stability for urea electrolysis. Up to date, although significant progress has been made on designing novel TM-LDHs nanosheets for UOR, the lowest overpotential value in recent reported publications is still larger than 900 mV at 10 mA cm$^{-2}$ (Table 3). Strategies to reduce the energy barrier of the *COO desorption step are required for accelerating UOR kinetics on TM-LDHs nanosheets. Therefore, new mechanisms and reaction pathways are urgently demanded to provide a guideline for constructing TM-LDHs nanosheets with optimal UOR performance.

**Table 3.** Recent reported TM-LDHs nanosheets for UOR.

| Materials | Electrolyte | Potential (V *vs.* RHE) | Current density (mA cm$^{-2}$) | Tafel (mV dec$^{-1}$) | Refs. |
|---|---|---|---|---|---|
| CoFeCr-LDHs | 0.33 M urea | 1.305 | 10 | 85 | [129] |
| NiMoV-LDHs | 0.33 M urea | 1.40 | 100 | 24.29 | [130] |
| FQD/CoNi-LDHs | 0.5 M urea | 1.36 | 10 | 17 | [131] |
| NiMn-LDHs | 0.5 M urea | 1.351 | 20 | 38.9 | [133] |
| NiCo-LDHs | 0.33 M urea | 1.36 | 50 | 45 | [134] |
| pa-NiFe-LDHs NS | 0.33 M urea | 1.362 | 30 | 33 | [135] |
| Cu-$Ni_3S_2$@NiFe-LDHs-200 | 0.5 M urea | 1.319 | 100 | 37.78 | [136] |
| NiFe-LDHs-F-4 | 0.33 M urea | 1.44 | 50 | 31 | [137] |
| NiAl-LDHs | 0.33 M urea | 1.42 | 10 | 59.8 | [30] |
| CoNi-LDHs-3 | 0.33 M urea | 1.39 | 100 | 16 | [138] |
| NiFe-LDHs/EG | 0.33 M urea | 1.34 | 50 | 44 | [139] |
| $Ce(OH)_3$@NiFe-LDHs | 0.5 M urea | 1.40 | 10 | 30.3 | [140] |
| H-NiFe-LDHs | 0.33 M urea | 1.397 | 10 | 41.7 | [141] |
| $MoS_2$/$Ni_3S_2$/NiFe-LDHs | 0.5 M urea | 1.396 | 100 | 36 | [142] |
| NiCo-LDHs- $[MoS_4]^{2-}$ | 0.33 M urea | 1.34 | 10 | 29 | [143] |
| CoMn-LDHs | 0.33 M urea | 1.326 | 10 | 73 | [144] |

## 4.2. Nitrogen Reduction Reaction

### *4.2.1. Mechanism*

Ammonia ($NH_3$) is not only widely used in the synthesis of fertilizers, drugs and chemicals, but also considered as an alternative fuel to store hydrogen.[145] Currently, the majority of $NH_3$ is produced through the conventional Haber-Bosch process at both high temperature and high pressure.[146] Under this circumstance, a possible alternative to the Haber-Bosch process is electrocatalytic NRR under mild conditions.[147] In acidic and alkaline environment, reaction equations of NRR are expressed as $N_2 + 6H^+ + 6e^- \rightarrow 2NH_3$ and $N_2 + 6H_2O + 6e^- \rightarrow 2NH_3 + 6OH^-$, respectively.[148]

Generally, NRR occurs through $N_2$ adsorption, hydrogenation, and subsequent $NH_3$ desorption.[149] Based on the order of cleavage and hydrogenation processes, dissociative and associative mechanism have been proposed.[150] For dissociative mechanism, the strong N≡N is cleaved prior to hydrogenation (Figure 12a). Because of extremely high bonding energy of N≡N, dissociative mechanism is mainly applicable for the industrial Haber-Bosch process with harsh reaction conditions. For associative mechanism, the cleavage and hydrogenation processes are preceded simultaneously. Considering the possible adsorption modes and hydrogenation sequences of nitrogen atoms, the associative mechanism is further classified into distal, alternating, and enzymatic pathways.[151] Figure 12b and Figure 12c depict the distal and alternating pathways. For both pathways, $N_2$ is initially end-on adsorbed on the catalyst surface. Afterwards, in the distal pathway, the nitrogen atom far away from the catalyst surface undergoes a preferential hydrogenation to release the first $NH_3$ molecule, and the remained nitrogen atom follows the same procedure to generate the second $NH_3$ molecule. Conversely, in the alternating pathway, two nitrogen atoms are hydrogenated alternately to produce $NH_3$ molecules after end-on adsorption on the catalyst surface. Lastly, in the enzymatic pathway (Figure 12d), two nitrogen atoms are firstly connected to the catalyst surface through side-on adsorption, and the subsequent hydrogenation process is same as the alternating pathway.

Recently, novel mechanism including Mars-van Krevelen (MvK) pathway [152] and surface-hydrogenation pathway [153] have been proposed through DFT calculations. Concretely, these two pathways are suitable for explaining NRR on the surface of transition metal nitrides and noble metals, respectively. In the MvK pathway (Figure 12e), lattice nitrogen atoms are firstly hydrogenated on the catalyst surface. Then, the newly supplemented $N_2$ are adsorbed to these resulting nitrogen vacancies and further hydrogenated to release $NH_3$. In the surface-hydrogenation pathway (Figure 12f), the potential-dependent Volmer reaction initially occurs to generate H* on the catalyst surface. Subsequently, surface H* activates $N_2$ to produce $N_2H_2$* with relatively high energy input. Finally, $NH_3$ is released by spontaneous reduction of $N_2H_2$*. Although new mechanistic pathways have validated to decrease the onset potential of NRR significantly, it is inevitably to ignore numerous practical factors (e.g. pH value) by relying on theoretical models.[154] Therefore, development of *in situ* characterization methods is essential to define the NRR process accurately.

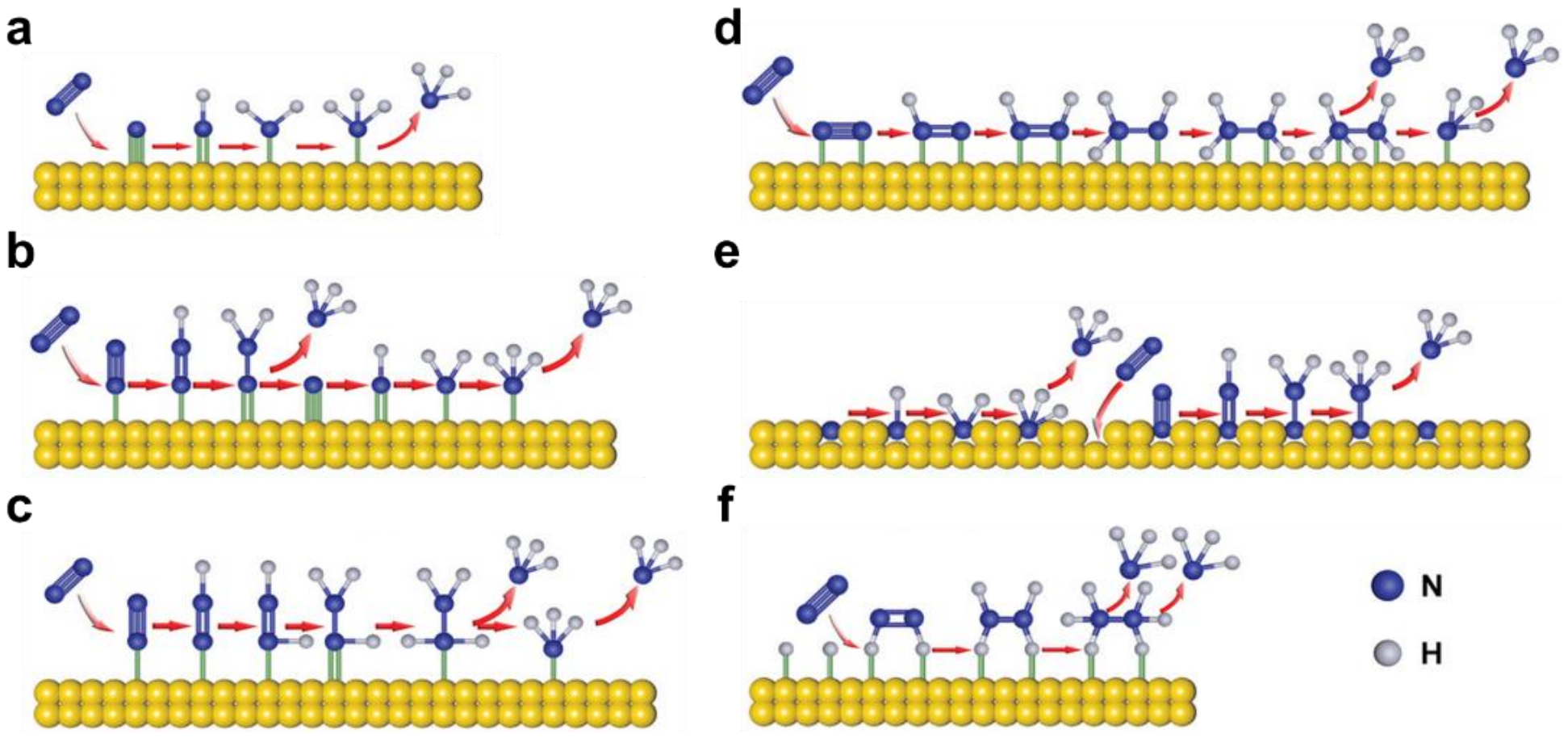

**Figure 12.** Schematic illustration of possible pathways for electrocatalytic conversion of $N_2$ to $NH_3$. Reproduced with permission.[149] Copyright 2022, Wiley.

*4.2.2. Research*

Compared with UOR, the application of TM-LDHs nanosheets for NRR has been less investigated. Zou *et al*. reported an efficient NRR electrocatalyst with the cross-link of NiCo-LDHs nanosheets.[155] By adjusting the Ni/Co ratio, the highest $NH_3$ yield was achieved as 52.8 μg $h^{-1}$ $mg_{cat}^{-2}$. The importance of Ni/Co ratio was further studied by *in situ* Raman measurements and DFT calculations. After comparing the activation energies of NRR process on four possible reaction sites, Co-Ni-Ni was proposed as the optimized active center for $NH_3$ production. Nevertheless, pristine TM-LDHs nanosheets are still subject to low conductivity, leading to sluggish $N_2$ activated/dissociated steps.

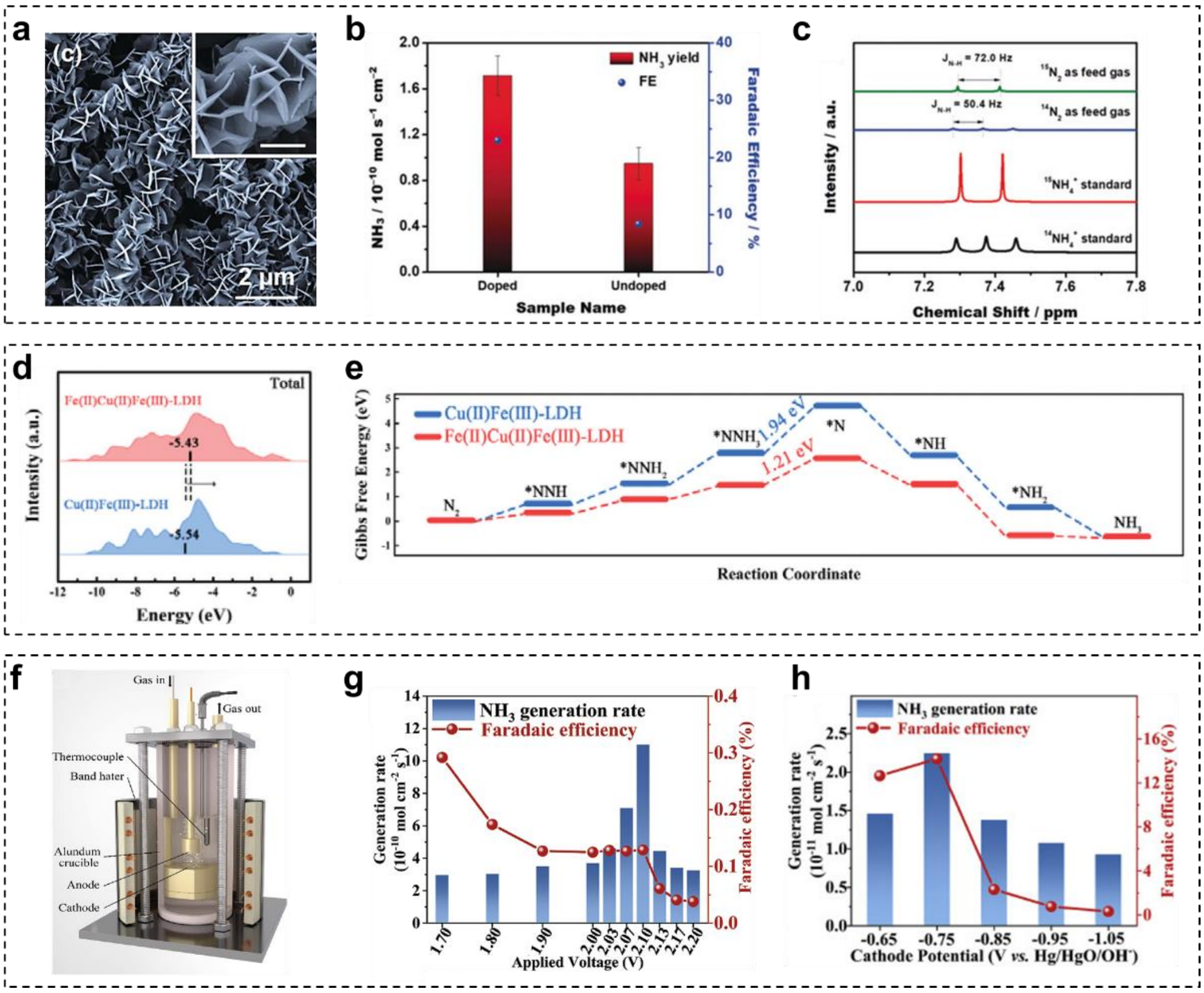

**Figure 13.** (a) SEM image of P-doped FeNi-LDHs nanosheets. (b) The comparison of $NH_3$ yields and FEs of undoped and P-doped FeNi-LDHs nanosheets at -0.5 V *vs.* RHE. (c) $^1$H NMR spectra of electrolytes with $^{15}N_2$ and $^{14}N_2$ as feed gas. Reproduced with permission.[156] Copyright 2020, Wiley. (d) The calculated d-band center of total elements and (e) NRR free-energy diagram for CuFe-LDHs and Fe(II)Cu(II)Fe(III)-LDHs nanosheets. Reproduced with permission.[157] Copyright 2022, Wiley. (f) Schematic diagram of the setup for single-chamber cell. $NH_3$ generation rates and FEs of the CoFe-LDHs nanosheets using (g) single-chamber cell and (h) H-cell. Reproduced with permission.[158] Copyright 2020, Elsevier.

Recently, several strategies have been validated to elevate NRR activity of TM-LDHs nanosheets, including doping heteroatoms (e.g., P and Se) and introducing low-valence transition metal ions. For example, Liu *et al*. coupled P-doped FeNi-LDHs nanosheet arrays with carbon nanofibers to achieve desired NRR performance.[156] As seen in Figure 13a, P-doped FeNi-LDHs nanosheets with the thickness of 20−30 nm were homogeneously distributed on the surface of carbon nanofibers. Attributed to the generation of extra empty d orbitals from P-doping, the maximum $NH_3$ yield and corresponding Faradaic efficiency (FE) were increased from $0.95 \times 10^{-10}$ to $1.7\times10^{-10}$ mol $s^{-1}$ $cm^{-2}$ and from 8.4% to 23%, respectively (Figure 13b).

Isotopic labeling experimental results in Figure 13c further demonstrated that the high selectivity toward $NH_3$ was originated from the feed $N_2$, not contaminants in the electrochemical testing system. Afterwards, Kong *et al.* applied a modified hydrothermal method to bring low-valence ferrous ion into CuFe-LDHs (Fe(II)Cu(II)Fe(III)-LDHs) nanosheets for effective NRR.[157] After adding the ferrous ions, the d-band center position was positively shifted from −5.54 to −5.43 eV (Figure 13d) and the energy barrier of potential-determining step was decreased from 1.94 to 1.21 eV (Figure 13e), suggesting the enhanced $N_2$ adsorption and reduction ability. As a result, Fe(II)Cu(II)Fe(III)-LDHs nanosheets realized a decent $NH_3$ yield rate of 33.1 ± 2.5 $\mu g\ h^{-1}\ mg_{cat}^{-1}$ with a FE of 21.7 ± 1.8% at −0.5 V *vs.* RHE, which outperformed that of pristine CuFe-LDHs nanosheets.

Apart from electronic structure modulation of TM-LDHs nanosheets, developing hierarchical heterostructure is apt to accelerate NRR kinetics by exposing sufficient active sites and facilitating the electronic/ionic transport. As reported by Arif *et al.*, the hierarchical CoVP@NiFeV-LDHs heterostructure showed superior NRR performance over the individual components (CoVP, NiFeV-LDHs nanosheets).[159] Due to highly durable CoVP substrate, efficient pH-universal $NH_3$ generation was also realized on this dedicated designed NRR electrocatalyst. Furthermore, proper selection of reaction system is another key point for fully reflecting NRR performance of TM-LDHs nanosheets. Qiao *et al.* evaluated NRR activities of CoFe-LDHs nanosheets using a single-chamber cell (Figure 13f) and a common H-cell separately.[158] Since HER and possible ammonia oxidation was inhibited to some extent in the H-cell, the FE of CoFe-LDHs nanosheets was elevated significantly after replacing single-chamber cell by H-cell (Figure 13g and Figure 13h). Nevertheless, the corresponding $NH_3$ generation capability in the H-cell was decreased dramatically. Thus, the balance between the highest $NH_3$ generation rate and FE should be carefully considered in the design of NRR experimental system. With tremendous efforts on TM-LDHs nanosheets, both the $NH_3$ generation rate and FE value have been greatly increased in the past few years. However, most recent reported TM-LDHs nanosheets with undesired selectivity (Table 4) cannot be compatible with the industrial requirements since the competing 2-electron HER is kinetically favored over 6-electron NRR.[160] Meanwhile, the overall $NH_3$ yield is also close related to the cleavage rate of strong N≡N on the surface of TM-LDHs nanosheets. Therefore, developing advanced NRR system with the composition of novel TM-LDHs nanosheets based electrocatalysts and non-aqueous electrolyte is expected to regulate the proton activity and realize efficient $NH_3$ synthesis.

**Table 4.** Recent reported TM-LDHs nanosheets for NRR.

| Materials | Electrolyte | FE | $NH_3$ Yield | Refs. |
|---|---|---|---|---|
| NiCo-LDHs | 0.1 M $Na_2SO_4$ | 11.5% | 52.8 μg $h^{-1}$ $mg_{cat}^{-2}$ | [155] |
| P-doped FeNi-LDHs | 0.1 M $Na_2SO_4$ | 23.0% | 1.72 × $10^{-10}$ mol $s^{-1}$ $cm^{-2}$ | [156] |
| Fe(II)Cu(II)Fe(III)-LDHs | 0.1 M $Na_2SO_4$ | 21.7% | 33.1 μg $h^{-1}$ $mg_{cat}^{-2}$ | [157] |
| CoVP@NiFeV-LDHs | 0.05 M $H_2SO_4$ | 13.8% | 1.6 μmol $h^{-1}$ $cm^{-2}$ | [159] |
| CoFe-LDHs | 25.6 mol $kg^{-1}$ KOH | 0.129% | 1.1 × $10^{-9}$ mol $s^{-1}$ $cm^{-2}$ | [158] |
| $Ni_{0.75}Fe_{0.25}Se_2$ | 0.1 M $Li_2SO_4$ | 12.3% | 5.64 μg $h^{-1}$ $cm_{cat}^{-2}$ | [161] |

## 5. Small Molecule Oxidation

### 5.1. Mechanism

Liquid small molecular fuels have been regarded as promising alternatives to gaseous hydrogen for fuel cells owing to their inexpensive price, high intrinsic energy density, and high safety.[162] Understanding the mechanism of small organic molecule electro-oxidation is essential for constructing efficient electrocatalysts. In this section, two representative reactions, methanol oxidation reaction (MOR) and ethanol oxidation reaction (EOR), are detailed elucidated.

MOR is a six-electron oxidaiton process to generate $CO_2$ as final product via *CO and *CO-free pathways. The *CO pathway in acidic and alkaline electrolytes are illustrated in Figures. 14a and 14b, respectively.[163] In the acidic electrolyte, MOR undergoes the following steps: $*CH_3OH \rightarrow *CH_2OH \rightarrow *CHOH \rightarrow *CHO \rightarrow *CO + H_2O \rightarrow *COOH \rightarrow * + CO_2$.[164] In the alkaline electrolyte, plenty $OH^-$ are participated in dehydrogenation steps: $*CH_3OH \rightarrow *CH_3O \rightarrow *CH_2O \rightarrow *CHO \rightarrow *CO + *OH \rightarrow * + CO_2$.[165] It is widely accepted that the MOR performance in alkaline electrolyte is better than that in the acidic electrolyte since a high concentration of $OH^-$ is beneficial for removing poisoning intermediate of *CO.[165] Alternatively, the key intermediate *CHO is converted to *HCOOH instead of *CO in the *CO-free pathway, and further hydrogenated into $CO_2$.[166]

In comparison with MOR, the mechanism of EOR is more complicated. As displayed in Figure 14c, EOR proceeds through dual pathways, which include a complete oxidation pathway and a partial oxidation pathway.[167] Concretely, the complete oxidation of ethanol leads to the production of $CO_2$ (acidic) or $HCO_3^-/CO_3^{2-}$ (alkaline) via the cleavage of the C-C bond and multiple dehydrogenation steps. However, the intrinsic slow kinetics of C-C bond splitting gives rise to the unfavorable occurrence of the complete oxidation pathway.[168] Thus, ethanol is prone to be partially oxidized to $CH_3COOH$ (acidic) or $CH_3COO^-$ (alkaline) via a four-electron transfer process. At present, mechanism of MOR and EOR are mostly proposed based on the noble metal electrocatalysts, such as Pt and Pd.[169] More attentions should be concentrated on the mechanistic study of earth-abundant transition-metal-based electrocatalysts.

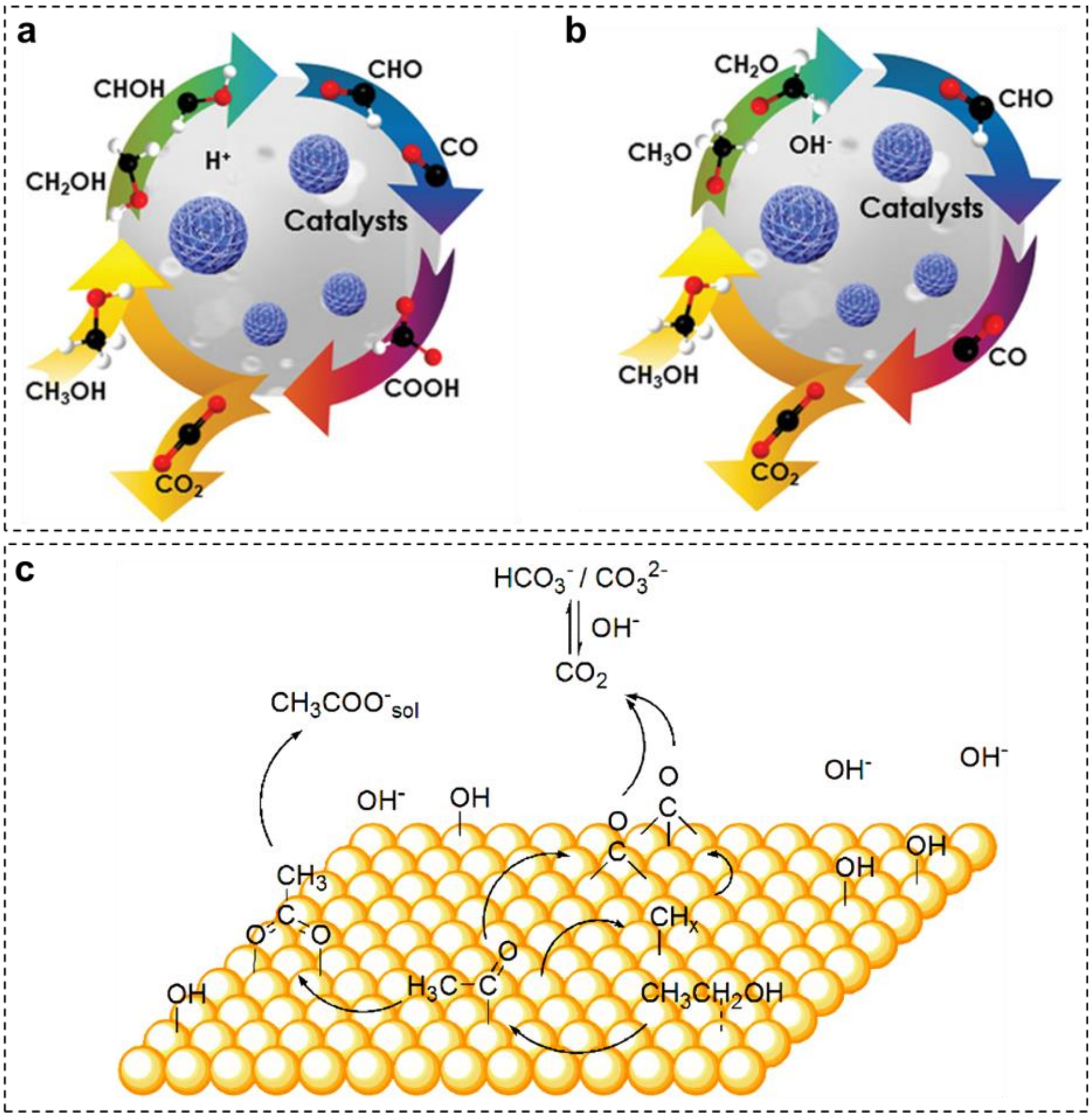

**Figure 14.** Schematic reaction pathways for MOR in (a) acidic and (b) alkaline electrolytes. Reproduced with permission.[163] Copyright 2021, Wiley. (c) Schematic reaction pathways for EOR in alkaline electrolyte. Reproduced with permission.[167] Copyright 2014, American Chemical Society.

### 5.2. Research

Considering low cost, ample abundance, and good compatibility with alkaline electrolyte, TM-LDHs nanosheets have drawn considerable attention as promising alternatives to traditional noble metal electrocatalysts in MOR and EOR. Initially, pristine TM-LDHs nanosheets were used without any modification, such as NiCo-LDHs [170] and NiAl-LDHs [18c]. Afterwards, the strategies related to the modulation of electronic configuration of TM-LDHs nanosheets have been performed to optimize the methanol and ethanol adsorption ability, leading to enhanced intrinsic MOR and EOR activities. For instance, Liu *et al.* prepared hierarchical $Cu_xCoCo$-LDHs nanosheets through the pseudomorphic transformation of $Cu_xCo$-MOFs.[171] Compared with the pristine CoCo-LDHs, all $Cu_xCoCo$-LDHs showed higher MOR activity and the lowest potential was achieved as 1.34 V *vs.* RHE to reach 50 mA $cm^{-2}$ (Figure 15a). In addition, the beneficial effect of Cu incorporation was identified by DFT calculations. According to Figures 15b and 15c, appropriate Cu incorporation gave rise to the ameliorated $CH_3OH$ adsorption and formic acid desorption abilities, which are essential for superior MOR activity. Very lately, the same group proposed and validated the pivotal role of hydroxyl vacancies in NiCo-LDHs nanosheets, which could also promote the adsorption of methanol. Thus, the monolayered NiCo-LDHs nanosheets with abundant hydroxyl vacancies achieved the desired potential of 1.35/1.40 V *vs.* RHE at 50/100 mA $cm^{-2}$.

Except for increasing the utilization of active sites, hybridization of TM-LDHs nanosheets with highly conductive materials is another mainstream research field for elevating MOR and EOR activities. For example, Li *et al.* reported a bifunctional hierarchical heterostructured $Co_xP$@NiCo-LDHs for efficient water/methanol electrolysis.[172] Owing to rapid electron transfer and highly exposed active sites, the current density towards MOR was obviously higher than that of counterpart materials (Figure 15d). Also, the heterostructure resulted in extremely high FEs (~100%) at various current densities (Figure 15e). Furthermore, $Co_xP$@NiCo-LDHs could be used as anodic and cathodic materials simultaneously in a symmetrical electrolyzer. As seen in Figure 15f, the voltage for hydrogen generation was reduced by 242 mV with the existence of methanol in alkaline electrolyte. Yang *et al.* constructed various NiFe-LDHs/carbon nanotube (CNT) composites with a classical hydrothermal method for efficient EOR.[173] Figure 15g clearly shows that NiFe-LDHs nanosheets were closely formed on the surface of spherical NiFe-LDHs/CNT composite. In comparison of CV and chronoamperometric analysis results (Figures 15h and 15i), the relatively high current density of NiFe-LDHs/CNT composite was originated from the positive effect of CNTs, which could accelerate charge transfer and provide robust mechanical strength. Albeit TM-LDHs nanosheets have been investigated to some extent as MOR/EOR electrocatalysts (Table 5), most existing researches related to TM-LDHs nanosheets remain at the stage of material design. Similar to UOR, the RDS of product desorption limits the whole kinetics of MOR/EOR. Thus, a comprehensive understanding of MOR/EOR mechanism on the surface of TM-LDHs nanosheets should be deciphered to design novel TM-LDHs nanosheets based electrocatalysts.

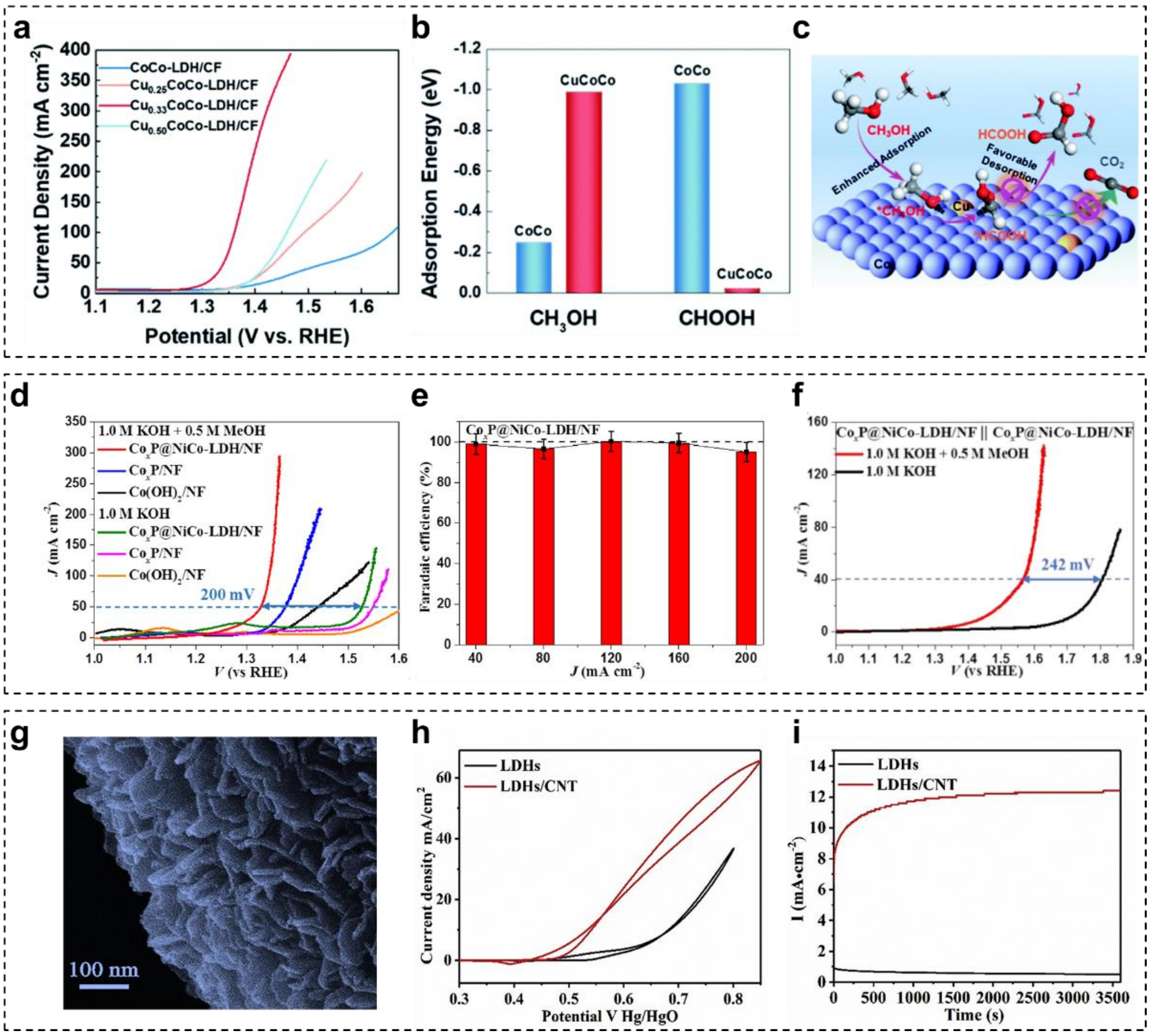

**Figure 15.** (a) LSV curves of $Cu_xCoCo$-LDHs/CFs in 1 M KOH + 3 M $CH_3OH$ electrolyte. (b) The calculated adsorption energy of $CH_3OH$ and formic acid molecules on CoCo-LDHs and CuCoCo-LDHs nanosheets. (c) The schematic process of MOR on the surface of CuCoCo-LDHs. Reproduced with permission.[174] Copyright 2022, Royal Society of Chemistry. (d) LSV curves of $Co(OH)_2$, $Co_xP$ and $Co_xP$@NiCo-LDHs towards MOR and OER. (e) FE of formate generation at different current densities. (f) LSV curves of $Co_xP$@NiCo-LDHs//$Co_xP$@NiCo-LDHs two-electrode electrolyzer with and without $CH_3OH$. Reproduced with permission.[172] Copyright 2020, Elsevier. (g) SEM image of NiFe-LDHs/CNT spherical composite. (h) CV

curves of NiFe-LDHs/CNT spherical composite and NiFe-LDHs powder sample in 1 M KOH + 1 M $CH_3CH_2OH$ electrolyte. (i) Chronoamperometric curves of NiFe-LDHs/CNT spherical composite and NiFe-LDHs powder sample. Reproduced with permission.[173] Copyright 2021, Elsevier.

**Table 5.** Recent reported TM-LDHs nanosheets for MOR/EOR.

| Materials | Methanol/Ethanol concentration | Potential (V *vs.* RHE) | Current density ($mA\ cm^{-2}$) | Refs. |
|---|---|---|---|---|
| $Ni_{0.33}Co_{0.67}(OH)_2$ | 0.5 M Methanol | 1.33 | 10 | [170] |
| $Cu_{0.33}$CoCo-LDHs | 3 M Methanol | 1.34 | 50 | [174] |
| NiCo-LDHs | 3 M Methanol | 1.35 | 50 | [175] |
| $Co_xP$@NiCo-LDHs | 0.5 M Methanol | 1.32 | 50 | [172] |
| $CeO_2$/NiV–LDHs | 0.5 M Methanol | 1.61 | 240 | [176] |
| NiFe-LDHs@$SnO_2$ | 0.5 M Methanol | 1.40 | 10 | [177] |
| $NiFe_xP$@NiCo-LDHs | 0.5 M Methanol | 1.40 | 50 | [178] |
| NiAl-LDHs | 1 M Ethanol | 1.59 | 45.8 | [18c] |
| NiFe-LDHs/CNT | 1 M Ethanol | 1.56 | 30.5 | [173] |

## 6. Biomass Derivatives Oxidation

Biomass is regarded as an ideal substitute for fossil resources to manufacture commercial industrial products.[179] Compared with traditional thermocatalytic technique, electrocatalysis possesses unique merits including benign operation condition, high efficiency, environmental friendliness, and the absence of additional reaction agents.[180] Motivated by these distinct advantages, tremendous research attention have been concentrated on electrocatalytic conversion of biomass derivatives to high-value-added products.[181] In this section, the application of TM-LDHs nanosheets for the electro-oxidation of specific biomass derivatives including 5-hydroxymethylfurfural (HMF), furfural (Fur)/furfural alcohol (FFA), glycerol, and benzyl alcohol are explicitly discussed.

### 6.1. 5-Hydroxymethylfurfural

In the last few decades, HMF has attracted great interest as one of top ten biobased chemicals by the US Department of Energy.[182] Since an aldehyde group and a hydroxymethyl group coexist in one HMF molecule, HMF can be further transformed into a number of valuable products, bridging the gap between the biomass resources and the chemical industry.[183] Among these important products, 2,5-furandicarboxylic acid (FDCA) have drawn extensive attention due to its usage in the production of poly(ethylene 2,5-furandicarboxylate) (PEF), a prospective substitute for petroleum-based polyethylene terephthalate (PET).[184]

Generally, HMF is oxidized to FDCA through two possible pathways, which are illustrated in Figure 16a.[185] The first step of path I and path II is the oxidation of the hydroxymethyl group and the aldehyde group of HMF to form 2,5-diformylfuran (DFF) and 5-hydroxymethyl-2-furancarboxylic acid (HMFCA) intermediates, respectively. Then both DFF and HMFCA intermediates in these two pathways are further oxidized to formyl-2-furancarboxylic acid (FFCA) intermediate and final product of FDCA. Depending on the adsorption ability of the hydroxymethyl group and the aldehyde group of HMF, HMF electro-oxidation process is highly dependent on pH value. Specifically, in a strong alkaline environment ($pH \geq 13$), the aldehyde group of HMF is more prone to be absorbed on the electrocatalyst surface in comparison of the hydroxymethyl group, resulting in HMFCA as the dominant intermediate. Conversely, in neutral or acidic environment ($pH < 13$), the hydration of the aldehyde group is inhibited, denoting that path I is a more favorable pathway

TM-LDHs nanosheets have been gradually deemed as efficient electrocatalysts for HMF electro-oxidation. Liu *et al.* first used NiFe-LDHs nanosheets to produce FDCA through electrochemical oxidation of HMF with both high yield (98%) and FE (99.4%).[186] Afterwards, Song *et al.* employed an electrochemical exfoliation method to construct oxygen deficient CoAl-LDHs nanosheet array (E-CoAl-LDH-NSA) for efficient electro-oxidation of HMF (Figure 16b).[187] Taking advantages of more exposed active sites and rich oxygen vacancies, the electrocatalytic performance of E-CoAl-LDHs-NSA toward HMF oxidation outperformed that of bulky CoAl-LDHs nanosheet array (B-CoAl-LDHs-NSA) (Figure 16c). According to the product analysis in Figure 16d, HMFCA and FFCA intermediates were observed during the electro-oxidation process, indicating that path II was preferentially occued on E-CoAl-LDH-NSA. More importantly, the beneficial role of oxygen vacancy was clarified by open circuit potential tests (Figure 16e) and free energy diagrams (Figure 16f). Compared with B-CoAl-LDHs-NSA, a larger OCP difference and a smaller energy barrier of the dehydrogenation step for E-CoAl-LDHs-NSA demonstrated that oxygen vacancies can enhance the adsorption of HMF and accelerate the kinetics of RDS, respectively, giving rise to high FDCA yield. In addition to binary LDHs, Zhang *et al.* reported ternary NiCoFe-LDHs nanosheets as efficient electrocatalysts for HMF oxidation.[188] The introduction of $Fe^{3+}$ into NiCo-LDHs nanosheets could reduce the interaction between metallic layers to decrease their thickness, and also modulate the electronic environment around active center. As a result, 95.5% HMF was converted to FDCA with a yield of 84.9% on NiCoFe-LDHs nanosheets within 1 h.

Apart from pristine TM-LDHs nanosheets, Xie *et al*. constructed CoFe@NiFe composites by coating NiFe-LDHs nanosheets on CoFe-LDHs nanosheets for HMF electro-oxidation.[189] Compared with the LSV curve of CoFe-LDHs nanosheets (Figure 16g), the CoFe@NiFe electrocatalyst exhibited better electrocatalytic activity due to their greater surface area, higher electronic conductivity, and faster reactant transport. Afterwards, within the potential range from 1.38 to 1.50 V *vs.* RHE, the charge for HMF conversion and OER gradually increased (Figure 16h), implying that OER was competitive with HMF oxidation at a high potential range. Thus, the optimal potential of HMF oxidation was chosen as 1.40 V *vs.* RHE and FDCA with 100% selectivity was realized at this potential with a FE of 99.8%. Furthermore, *in situ* Raman spectra at different potentials (Figure 16i) revealed that $Ni^{3+}$ was the main active species during HMF electro-oxidation.

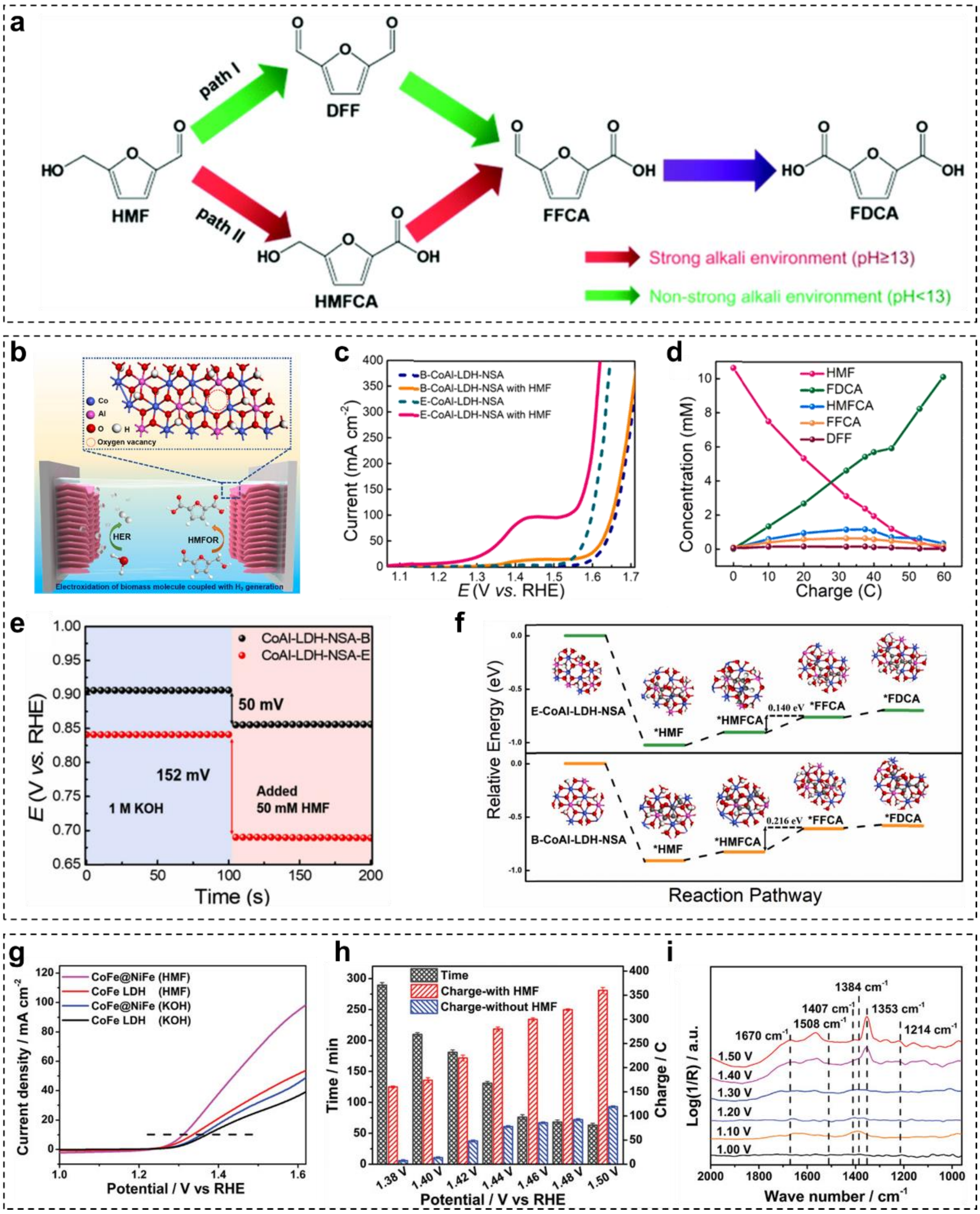

**Figure 16.** (a) Two pathways of electrochemical oxidation of HMF. Reproduced with permission.[185a] Copyright 2021, Royal Society of Chemistry. (b) Illustration of E-CoAl-LDH-NSA for electro-oxidation of HMF coupled with hydrogen generation. (c) LSV curves of B-CoAl-LDH-NSA and E-CoAl-LDH-NSA in 1 M KOH with and without 10 mM HMF (d) Concentration of HMF and corresponding oxidation products with increasing charge for E-CoAl-LDH-NSA. (e) Open circuit potential comparison of B-CoAl-LDHs-NSA and E-CoAl-LDHs-NSA with the addition of 50 mM HMF after 100 s. (f) Free energy diagrams of HMF electro-oxidation to FDCA on E-CoAl-LDH-NSA and B-CoAl-LDH-NSA. Reproduced with permission.[187] Copyright 2021, Elsevier. (g) LSV curves of CoFe@NiFe and CoFe-LDHs nanosheets in 1 M KOH with and without 10 mM HMF. (h) Comparison of the total time for HMF conversion, and the consumed charge with and without HMF at different potentials. (i) *In situ* Raman spectra of CoFe@NiFe electrocatalyst at different potentials during HMF electro-oxidation. Reproduced with permission.[189] Copyright 2021, Wiley.

### 6.2. Furfural/Furfuryl Alcohol

Fur is a lucrative hemicellulose-derived substance for various applications, such as pharmaceuticals, oil refining, and pesticides.[190] The traditional route for Fur oxidation is proceeded at both high temperature and high-pressure oxygen using oxidants or noble metal catalysts (e.g., Pt, Au, Pd, Ru).[191] Alternatively, Fur electrochemical oxidation is an attractive replacement to the conventional thermocatalytic technique since the new oxidation process can be operated in normal pressure and temperature without adding external oxygen.[192] As seen in Figure 17a, Fur could be electro-oxidized to furoic acid (FurAc) by converting the aldehyde group to carboxylic acid.[193] However, Cannizzaro reaction is always proceeded in Fur electro-oxidation process under alkaline condition, leading to a undesired FurAc yield (< 50%).[194] Therefore, further oxidation of another product in Cannizzaro reaction, FFA, becomes important for increasing FurAc yield. Figure 17b clearly depicts the pathway of FFA oxidation.[195]

Very recently, TM-LDHs nanosheets are reported as the efficient electrocatalysts for FurAc generation. For example, Fan *et al.* assembled NiFe-LDHs nanosheets on $Ni_xSe_y$ nanowire arrays to construct core-shell $Ni_xSe_y$-NiFe-LDHs electrocatalyst for Fur electro-oxidation.[196] As displayed in Figure 17c, the conversion of Fur was complete after passing 116 C to obtain FurAc yield of 99.7% and FE of 99.5% on this core-shell electrocatalyst. Afterwards, the FurAc yield and FE exhibited inconspicuous degradation after six cycles, demonstrating robust stability (Figure 17d). Herein, the satisfied electrocatalytic performance for FF electro-oxidation was attributed to the following advantages (Figure 17e): the highly conductive $Ni_xSe_y$ nanowires facilitated electron transport and provided robust mechanical support; NiFe-LDHs nanosheets with a large surface area offered sufficient active sites; and the open structure ensured electrolyte penetration. Liu *et al.* synthesized 3D hierarchical NiFe microflowers as the efficient electrocatalyst via electrodeposition strategy.[195] As depicted in Figure 17f, the selective oxidation of FFA occurred in advance of OER. By adjusting preparation and testing parameters, optimal NiFe-LDHs nanosheets achieved the highest FFA conversion of 81.4% and FurAc selectivity of 94.0% within 3 h (Figures 17g and Figure 17h).

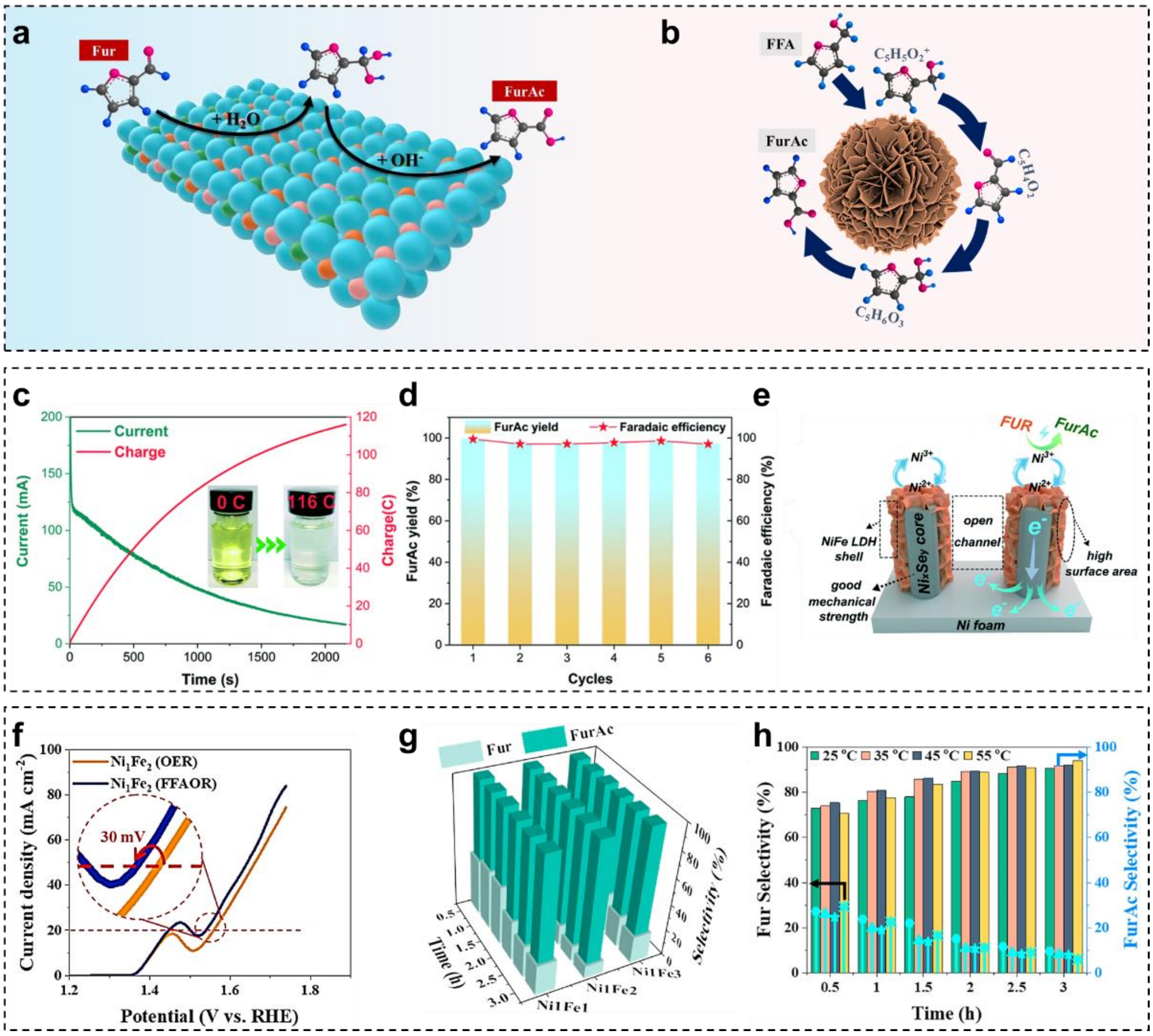

**Figure 17.** The pathway of electro-oxidation of (a) Fur to FurAc. Reproduced with permission.[193] Copyright 2021, Royal Society of Chemistry. (b) FFA to FurAc. Reproduced with permission.[195] Copyright 2021, KeAi Publishing Communications Ltd. (c) Current-time curve of Fur oxidation on the $Ni_xSe_y$-NiFe-LDHs electrocatalyst under controlled potential. (d) FurAc yield and FE of the $Ni_xSe_y$-NiFe-LDHs electrocatalyst after six cycles. (e) The mechanism proposed for Fur electro-oxidation on the $Ni_xSe_y$-NiFe-LDHs electrocatalyst. Reproduced with permission.[196] Copyright 2022, Royal Society of Chemistry. (f) LSV curves of NiFe-LDHs nanosheets in 1 M KOH with and without FFA. The selectivity of products on NiFe-LDHs nanosheets with (g) different $Ni^{2+}/Fe^{3+}$ ratio and (h) different reaction temperature. Reproduced with permission.[195] Copyright 2021, KeAi Publishing Communications Ltd.

### 6.3. Glycerol

Glycerol, a C3 platform molecule, is the main by-product from biodiesel production.[197] Numerous high-value-added products could be obtained by selective oxidation of glycerol, including dihydroxyacetone, glyceric acid, tartronic acid, mesoxalic acid, glycolic acid, and formic acid.[198] Unlike traditional thermocatalytic glycerol oxidation reactions (GOR), electrocatalytic GOR is a promising route, which can proceed under moderate temperature and atmospheric pressure with low cost and high efficiency.[199] Electrocatalytic GOR is a multi-electron transfer process, which is generally classified into two reaction routes. As described in Figure 18a, the rightward direction and downward direction correspond to the cleavage of C-C bond and oxidation of alcohol to aldehyde or carboxylic acid through PCET, respectively. 1) oxidation of alcohol into aldehyde or carboxylic acid through PCET (downward direction).[200] For formic acid production, the theoretical FE of one pathway with glyceraldehyde and glycolaldehyde intermediates is achieved as 100%, whereas maximum FE of other pathways is 66.6%.

Noble metals are most common electrocatalysts for GOR.[201] Lately, TM-LDHs nanosheets have been utilized as low-cost replacements to these precious electrocatalysts for GOR. For example, Dong *et al*. developed an electrochemical treatment on NiV-LDHs nanosheets to generate plenty of Ni(III) active sites for efficient GOR (Figure 18b).[202] Electrochemically regulated NiV-LDHs (E-NiV-LDHs) nanosheets presented superior GOR activity with a low potential of 1.23 V *vs.* RHE to drive 10 mA $cm^{-2}$ (Figure 18c). After optimizing the potential for GOR, high glycerol conversion of 98% and FE of 94% were demonstrated (Figure 18d). Subsequently, GOR mechanism of E-NiV-LDHs nanosheets was further investigated by *in situ* techniques. *In situ* Fourier transform infrared (FTIR) and Raman spectra in Figure 18e and Figure 18f revealed that the final product formate should be formed during GOR process and Ni(III) active sites were responsible for high-efficient formate generation, respectively. Similarly, NiCo-LDHs nanosheets were also validated as the efficient GOR electrocatalysts.[203] Intriguingly, it is found that the formation of oxygen vacancies on the surface of NiCo-LDHs nanosheets during GOR was conducive to the cleavage of the 2nd C-C bond. Therefore, NiCo-LDHs nanosheets displayed excellent GOR activity to afford 100 mA $cm^{-2}$ at 1.35 V *vs.* RHE with formate selectivity of 94.3%.

### 6.4. Benzyl Alcohol

Benzaldehyde and benzoic acid are important high-value-added chemicals that are extensively employed in pharmaceutical, antiseptic, and agrochemical industries.[204] Conventional industrial synthesis of benzaldehyde and benzoic acid from benzyl alcohol always involve the utilization of toxic reagents under harsh condition.[205] Therefore, it is extremely desired to develop a new sustainable catalytic oxidation method for benzyl alcohol. Benzyl alcohol electro-oxidation is an emerging technique with the benefits of benign operation conditions and environmental friendliness.[206] A classical pathway of benzyl alcohol electro-oxidation is shown in Figure 18g.[207] The theoretical potential of benzyl alcohol electro-oxidation and OER are −0.48 and −1.23 V, respectively, revealing that benzyl alcohol electro-oxidation reaction is more advantageous over OER in the thermodynamic aspect. Nevertheless, most electrocatalysts for benzyl alcohol electro-oxidation are suffered from insufficient current density, which cannot meet the industrial requirement.[208]

Recently, amorphous NiCo-LDHs nanosheets was used as brilliant high-performance electrocatalyst for the electro-oxidation of benzyl alcohol to benzoic acid.[207] As displayed in Figure 18h, the as-prepared NiCo-LDHs nanosheets realized an ultra-large current density over 400 mA $cm^{-2}$ before the appearance of OER. At the maximum potential of 1.5 V *vs.* RHE, 99% benzyl alcohol was converted to benzoic acid with a high yield of approximate 99% within 15 min (Figure 18i). Furthermore, *in situ* Raman measurements were carried out on NiCo-LDHs nanosheets to decipher the actual active center during electro-oxidation process. The corresponding reaction cell was designed as Figure 18j. As the anodic potential was cycled between 0.2 and 0.5 V *vs.* Ag/AgCl, the occurrence of characteristic peaks at 466 and 542 $cm^{-1}$ confirmed that NiOOH was the active species in the benzyl alcohol oxidation reaction (Figure 18k). This work sheds the light on designing the electrocatalysts for biomass derived oxidation with industrial level.

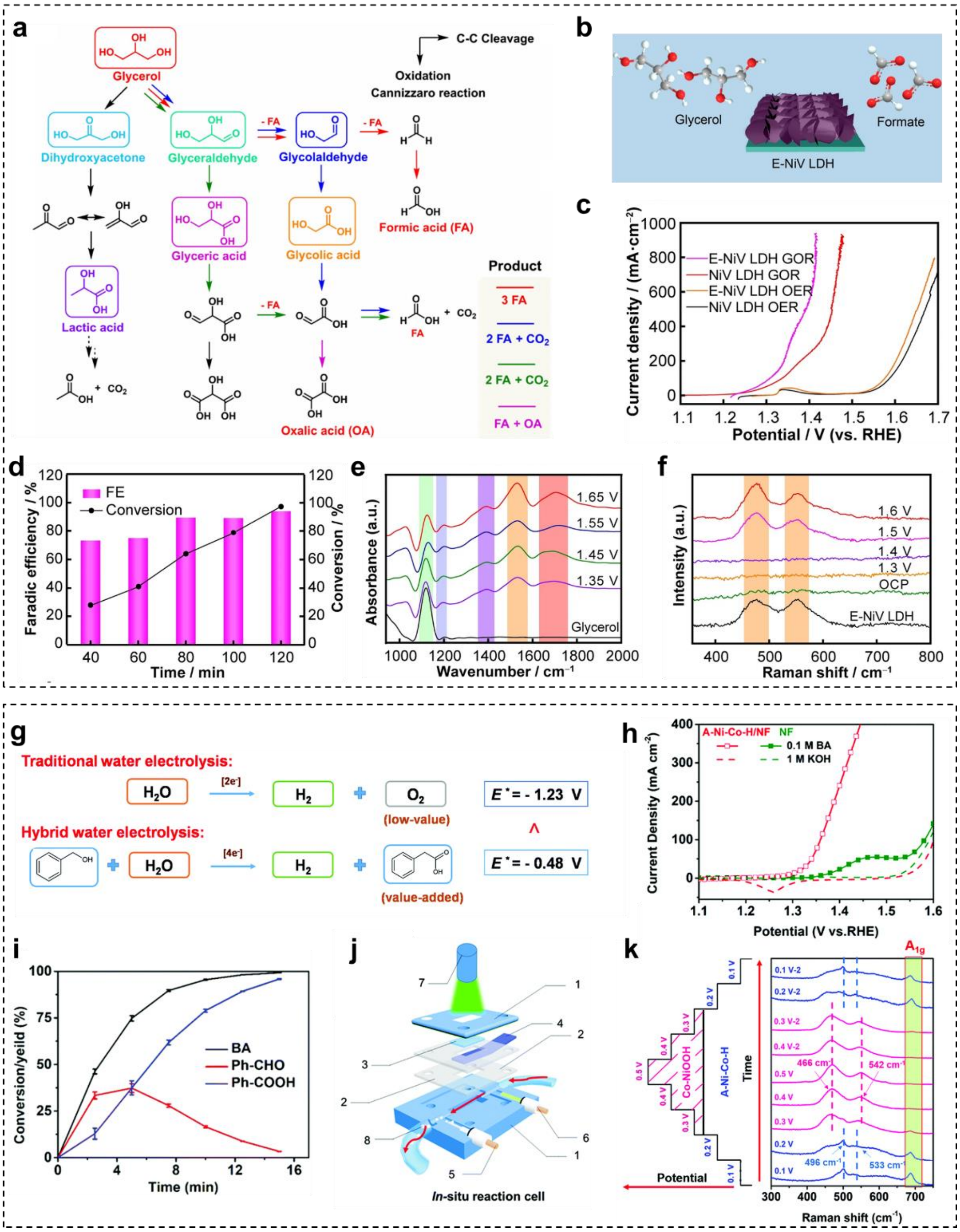

**Figure 18.** (a) Reaction pathways for the electro-oxidation of glycerol. Reproduced with permission.[200] Copyright 2022, Wiley. (b) Scheme of E-NiV-LDHs nanosheets for production of formate. (c) LSV curves of E-NiV-LDHs and NiV-LDHs nanosheets in 1 M KOH with and without 0.1 M glycerol. (d) Time-dependent glycerol conversion and FE on E-NiV-LDHs nanosheets at 1.46 V *vs.* RHE. (e) *In situ* FTIR spectra at different potentials during GOR. (f) *In situ* Raman spectra of E-NiV-LDHs nanosheets at different potentials during GOR. Reproduced with permission.[202] Copyright 2022, Nonferrous Metals Society of China. (g) Reaction pathways for the electro-oxidation of benzyl alcohol. (h) LSV curves of NiCo-LDHs

nanosheets in 1 M KOH with and without 0.1 M benzyl alcohol at 10 mV $s^{-1}$. (i) Time-dependent conversion/yield in benzyl alcohol electro-oxidation reaction at a potential of 1.5 V *vs.* RHE. (j) The schematic illustration of the *in situ* reaction cell. (k) *In situ* Raman spectra of NiCo-LDHs nanosheets under different potentials. Reproduced with permission.[207] Copyright 2020, Royal Society of Chemistry.

Summarily, TM-LDHs nanosheets can be regarded as one of the most active materials for biomass derivatives oxidation. Table 6 lists recent reported TM-LDHs nanosheets as efficient electrocatalysts for various biomass derivatives oxidation. However, several problems still need to be resolved. The first problem is to uncover the active center of TM-LDHs nanosheets more distinctly. The second problem is the limited stability of TM-LDHs nanosheets with the prolonged exposure to strong alkaline electrolyte. Hence, in order to further enhance electrocatalytic performance of TM-LDHs nanosheets for biomass derivatives oxidation, new strategies should be developed in both experimental and theoretical aspects. Furthermore, since *OH intermediate is involved in electrocatalytic oxidation of biomass derivatives,[209] the binding energy between *OH and TM-LDHs nanosheets should be optimized in the future material design.

**Table 6.** Recent reported TM-LDHs nanosheets for biomass oxidations.

| Materials | Electrolyte | Yield (%) | FE [%] @ E [V *vs.* RHE] | Ref. |
|---|---|---|---|---|
| NiFe-LDHs | 1 M KOH/10 mM HMF | FDCA, 98.0 | 99.4 @ 1.23 V | [186] |
| CoAl-LDHs | 1 M KOH/10 mM HMF | FDCA, / | 99.4 @ 1.52 V | [187] |
| d-NiFe-LDHs | 1 M KOH/10 mM HMF | FDCA, 96.8 | 84.47 @ 1.48 V | [210] |
| NiCoFe-LDHs | 1 M NaOH/10 mM HMF | FDCA, 84.9 | ~90 @ 1.52 V | [188] |
| NiCoMn-LDHs | 1 M NaOH/1 mM HMF | FDCA, 91.7 | ~65 @ 1.50 V | [193] |
| $Cu_xS$@NiCo-LDHs | 1 M KOH/10 mM HMF | FDCA, ~99.0 | 99.0 @ 1.32 V | [211] |
| CoFe@NiFe | 1 M KOH/10 mM HMF | FDCA, 99.8 | 99.8 @ 1.40 V | [189] |
| NiCoMn-LDHs | 1 M NaOH/1 mM Fur | FurAc, 92.4 | ~70 @ 1.50 V | [193] |
| $Ni_xSe_y$-NiFe LDHs | 1 M KOH/20 mM Fur | FurAc, 99.7 | 99.5 @ 1.423 V | [196] |
| NiFe-LDHs | 1 M KOH/1 mM FFA | FurAc, / | 90.4 @ 1.60 V | [195] |
| E-NiV-LDHs | 1 M KOH/0.1 M glycerol | Formate, 94.0 | 94 @ 1.46 V | [202] |
| NiCo-LDHs | 1 M KOH/0.1 M glycerol | Formate, 94.3 | ~100 @ 1.474 V | [203] |
| A-NiCo-LDHs | 1 M KOH/0.1 M benzyl alcohol | Benzyl alcohol, 99.0 | 93.5 @ 1.50 V | [207] |

## 7. Summary and Perspective

In this review, we have discussed the synthetic strategies in preparing TM-LDHs nanosheets, and summarized the recent advances of TM-LDHs nanosheets for various electrocatalytic applications including OER, HER, UOR, NRR, small molecule oxidation and biomass derivatives oxidation in details. Among these electrocatalytic applications, the mechanism and performance optimization of OER, HER, and UOR using TM-LDHs nanosheets have been thoroughly studied. Although TM-LDHs nanosheets have drawn tremendous attention owing to their low cost, easy accessibility, and changeable intrinsic electronic structure, there is still more space for performance improvements. Hence, the following perspectives are proposed:

1) Scale-up: Developing cost-effective fabrication methods of large-scale TM-LDHs nanosheets based electrodes should be a prerequisite for advancing toward industrialization. Nevertheless, most current fabrication processes are restricted to the lab-scale level. Few researchers have conducted fundamental studies on the architecture of electrodes with industrial-level size. Hence, the design criteria of TM-LDHs nanosheets from the lab to industry should be better understood.

2) Substrate: Due to high surface energy, aggregation of TM-LDHs nanosheets unavoidably occurs without using any substrate, leading to a slow electron transfer, limited contract between the electrolyte and electrocatalyst, and suppressed electrocatalytic activity of active sites. There is an emerging trend to prepare self-supporting electrocatalysts by *in situ* growing TM-LDHs nanosheets on highly conductive 3D substrates to improve conductivity and stability. Nowadays, the main types of substrates are NF, cobalt foam, and carbon paper. For maximizing the excellent performance of TM-LDHs nanosheets, it is also essential to build a substrate with a quick desorption rate for products.

3) Surface reconstruction: Intensive surface reconstruction has been identified on TM-LDHs nanosheets during electrocatalytic process, resulting in true active center for the enhanced electrocatalytic activity. However, the fundamental origin of the self-reconstruction of TM-LDHs nanosheets and the factors governing the relationship between the self-reconstruction and electrocatalytic performance are not fully unveiled. Thus, to better design TM-LDHs nanosheets based electrocatalysts, *in situ*/operando characterization techniques (e.g., electrochemistry or microscopy/spectroscopy) are required to uncover the mechanism of surface reconstruction at the molecular level.

4) Long-term stability in practical ambient environments: Currently, most electrocatalytic research using TM-LDHs nanosheets are proceeded in alkaline electrolyte with high purity. However, strong alkaline condition causes severe corrosion of electrodes and other corresponding components for a long-term run. Also, various impurities, such as dust, and corrosive ions in the actual water could deactivate active sites and further corrode the TM-LDHs nanosheets. Thus, it is highly desired to elucidate the electrocatalytic mechanism of TM-LDHs nanosheets based electrocatalysts in neutral electrolyte solution and modify the surface of TM-LDHs nanosheets to mitigate the corrosion in actual water.

With this comprehensive overview, we believe that the readers could follow the current trends in designing and evaluating TM-LDHs nanosheets for electrocatalytic applications. We hope this review can provide inspiration to promote the development and application of related high-performance electrocatalysts.

**Conflict of Interest**

The authors declare no conflict of interest.

**Acknowledgments**

This work was supported by National Natural Science Foundation of China (No. 22078374), National Key R&D Program of China (No. 2020YFC1807600), National Ten Thousand Talent Plan, the Scientific and Technological Planning Project of Guangzhou (No. 202206010145), China, and JSPS KAKENHI Grant 22H01855, Japan.

Received: ((will be filled in by the editorial staff))

Revised: ((will be filled in by the editorial staff))

Published online: ((will be filled in by the editorial staff))

**Author Biographies and Photographs**

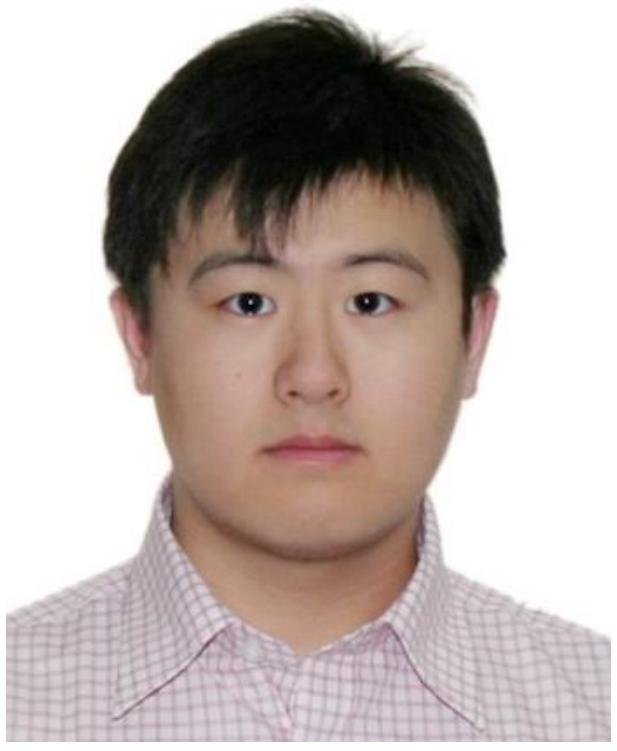

**Yuchen Wang** received his Ph.D. degree in Mechanical Engineering from the University of Miami in 2017. He is currently a postdoc research associate at the School of Environmental Science and Engineering, Sun Yat-sen University. His present research interests focus on nanomaterials in energy storage and conversion applications.

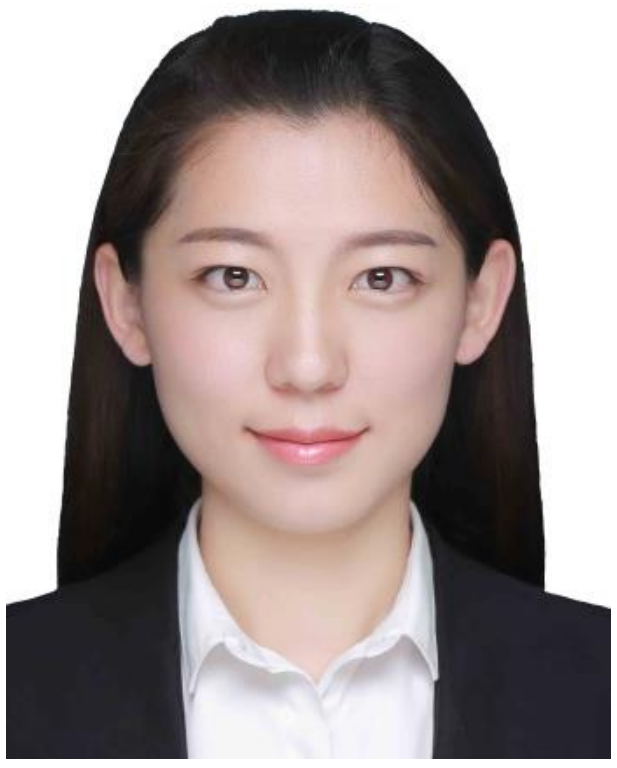

**Man Zhang** is currently pursuing her Ph.D. degree under the supervision of Prof. Kai Yan in the School of Environmental Science and Engineering, Sun Yat-sen University. Her current research interests focus on nanomaterials for electrocatalytic applications, such as oxygen evolution reaction, hydrogen evolution reaction, biomass derivative conversion.

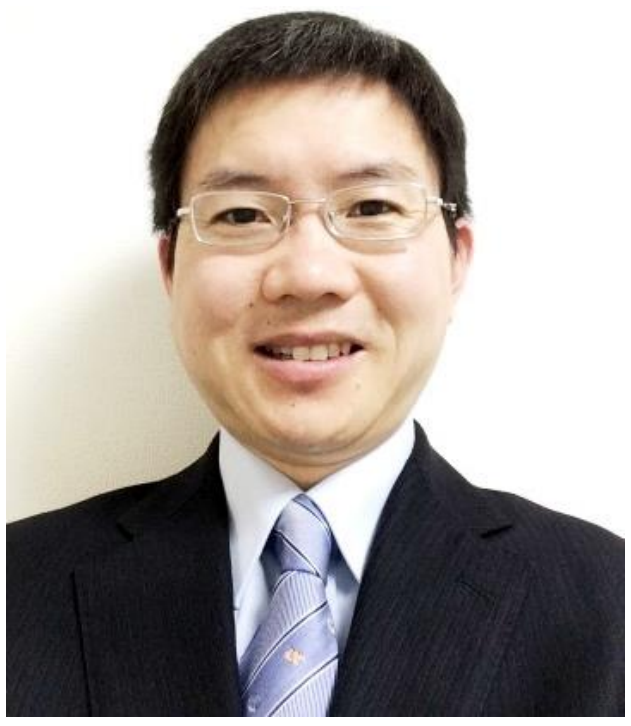

**Guoqing Guan** is a full professor of Hirosaki University, Japan. He received his Ph.D. degree in Chemical Engineering from Sichuan University in 1995. He also worked as an Alexander von Humboldt research fellow in Germany and a JSPS research fellow in Japan. His research interests are water splitting for hydrogen production, electrochemical reduction of $CO_2$ and biomass derivatives oxidation for high-value-added chemicals production, energy materials for energy conversion, coal/biomass pyrolysis and gasification, bio-refinery, and heterogeneous catalysts.

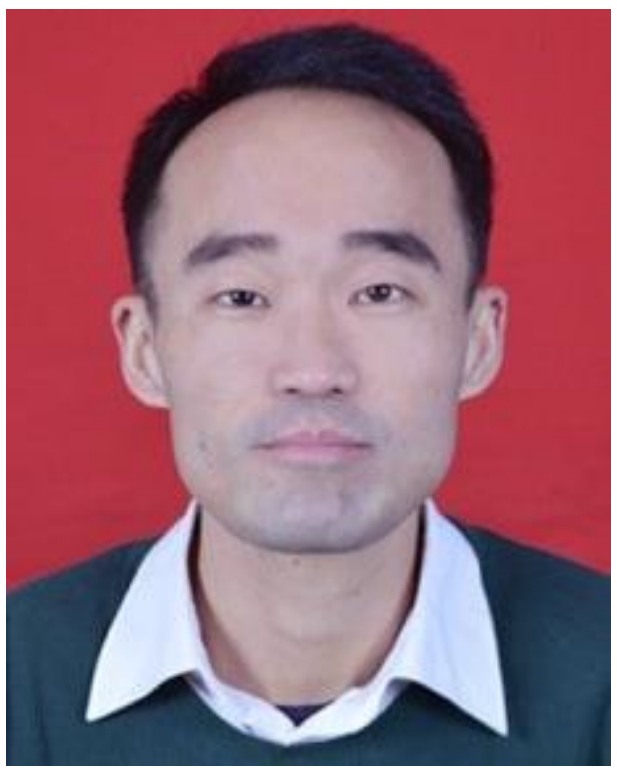

**Kai Yan** is a full professor at the School of Environmental Science and Engineering, Sun Yat-sen University. He received his Ph.D. degree from Max-Planck-Institute for Coal Research and RWTH Aachen University in 2011. Then he obtained an Ontario Government Postdoctoral Fellowship at Lakehead University (2012-2013) and joined Brown University as a postdoctoral research associate (2013-2016). His current interests include the synthesis of nanostructured materials for clean energy and environment-related applications.

Current fabrication strategies to design TM-LDHs nanosheets are summarized. The electrocatalytic applications of these as-fabricated TM-LDHs nanosheets in oxygen evolution reaction, hydrogen evolution reaction, urea oxidation reaction, nitrogen reduction reaction, small molecule oxidation and biomass derivatives upgrading are articulated through systematically discussing the corresponding fundamental design principles and reaction mechanism.

Yuchen Wang,† Man Zhang,† Yaoyu Liu, Zhikeng Zheng, Biying Liu, Guoqing Guan,* Kai Yan*

**Recent Advances in Transition-Metal-Based Layered Double Hydroxides Nanosheets for Electrocatalytic Energy Conversion**

ToC figure

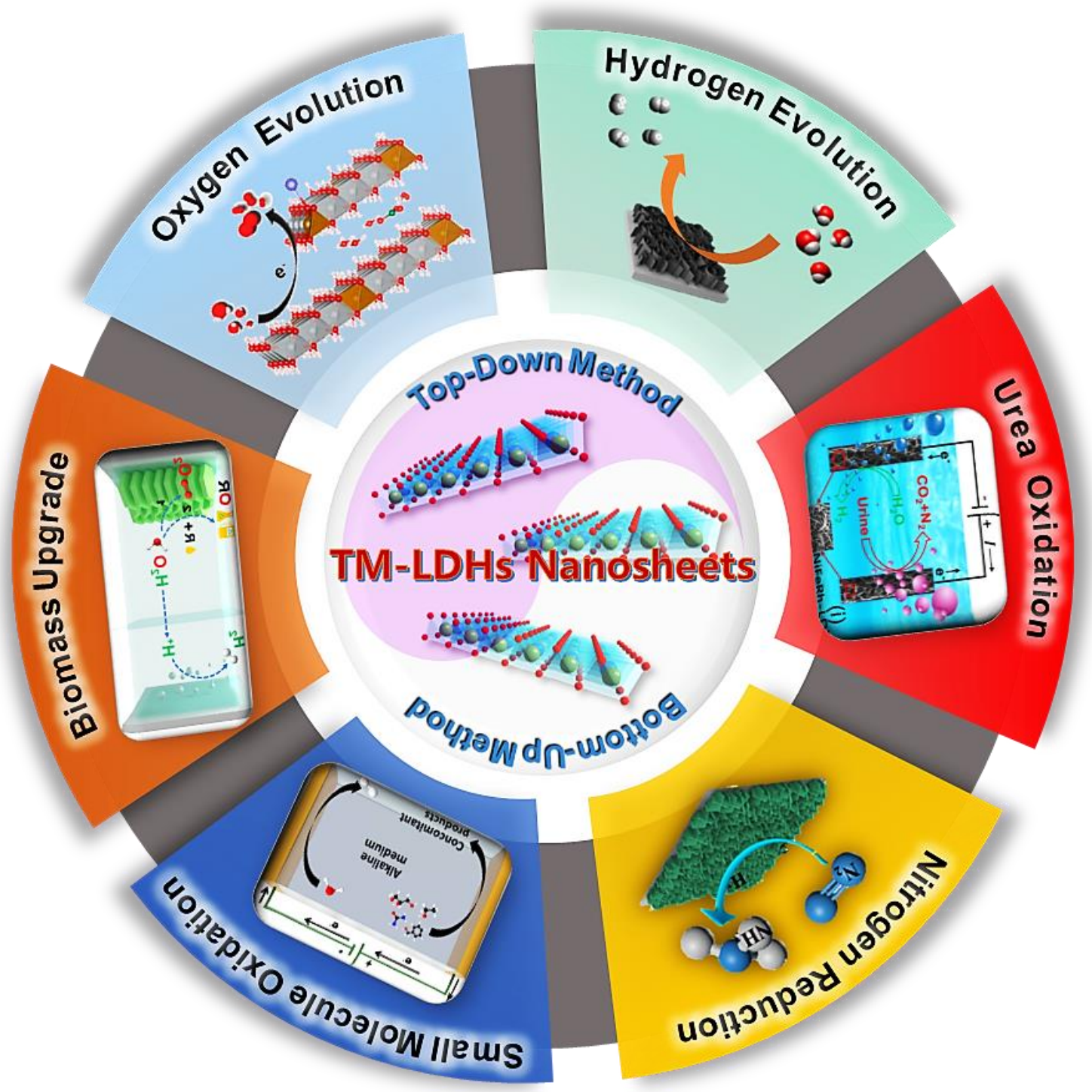